\documentclass[nonacm,acmsmall,screen]{acmart}
\AtBeginDocument{%
  }

\setcopyright{none}
\copyrightyear{xxxx}
\acmYear{xxxx}
\acmDOI{XXXXXXX.XXXXXXX}

\acmJournal{TOMS}
\acmVolume{xx}
\acmNumber{x}
\acmArticle{xxx}
\acmMonth{1}

\setcitestyle{nosort}

\usepackage[cachedir=images/cache]{graphicscache}
\usepackage{graphicx}
\usepackage{color}
\usepackage{bm}
\usepackage{enumitem}
\usepackage{algorithm}
\usepackage[noend]{algpseudocode}
\usepackage{hyperref}
\newcommand{\DFT}{\mathrm{DFT}}
\newcommand{\FFT}{\mathrm{FFT}}
\newcommand{\NTT}{\mathrm{NTT}}

\newcommand{\fl}{\,\mathbf{fl}\,}

\usepackage{cleveref}

\crefname{figure}{Figure}{Figures}
\Crefname{figure}{Figure}{Figures}
\crefname{equation}{}{}
\Crefname{equation}{}{}
\crefname{algorithm}{Algorithm}{Algorithms}
\Crefname{algorithm}{Algorithm}{Algorithms}
\crefname{table}{Table}{Tables}
\Crefname{table}{Table}{Tables}

\renewcommand{\Re}{\operatorname{Re}}
\renewcommand{\Im}{\operatorname{Im}}

\usepackage{autonum}
\begin{document}

%%
%% The "title" command has an optional parameter,
%% allowing the author to define a "short title" to be used in page headers.
\title{Computing FFTs at Target Precision Using Lower-Precision FFTs}

%%
%% The "author" command and its associated commands are used to define
%% the authors and their affiliations.
%% Of note is the shared affiliation of the first two authors, and the
%% "authornote" and "authornotemark" commands
%% used to denote shared contribution to the research.
\author{Shota Kawakami}
\email{kawakami@hpcs.cs.tsukuba.ac.jp}
\orcid{0009-0000-7062-2874}
\affiliation{%
  \institution{Graduate School of Science and Technology, University of Tsukuba}
  \streetaddress{1-1-1 Tennodai}
  \city{Tsukuba}
  \state{Ibaraki}
  \postcode{305-8577}
  \country{Japan}
}

\author{Daisuke Takahashi}
\email{daisuke@cs.tsukuba.ac.jp}
\orcid{0000-0003-1357-5770}
\affiliation{%
  \institution{Center of Computational Sciences, University of Tsukuba}
  \streetaddress{1-1-1 Tennodai}
  \city{Tsukuba}
  \state{Ibaraki}
  \postcode{305-8573}
  \country{Japan}
}

%%
%% By default, the full list of authors will be used in the page
%% headers. Often, this list is too long, and will overlap
%% other information printed in the page headers. This command allows
%% the author to define a more concise list
%% of authors' names for this purpose.
\renewcommand{\shortauthors}{Kawakami and Takahashi}

%%
%% The abstract is a short summary of the work to be presented in the
%% article.
\begin{abstract}
Modern processors deliver higher throughput for lower-precision arithmetic than for higher-precision arithmetic.
For matrix multiplication, the Ozaki scheme exploits this performance gap by splitting the inputs into lower-precision components and delegating the computation to optimized lower-precision routines.
However, no similar approach exists for the fast Fourier transform (FFT).
Here, we propose a method that computes target-precision FFTs using lower-precision FFTs by applying the Ozaki scheme to the cyclic convolution in the Bluestein FFT.
The split component convolutions are computed exactly using the number theoretic transform (NTT), an FFT over a finite field, instead of floating-point FFTs, combined with the Chinese remainder theorem.
We introduce an upper bound on the number of splits and an NTT-domain accumulation strategy to reduce the NTT call count.
As a concrete implementation, we implement a double-precision FFT using 32-bit NTTs and confirm reduced relative error compared with those for FFTs based on FFTW and Triple-Single precision arithmetic, with stable error across FFT lengths, at most 96 NTT calls, or 64 NTT calls with NTT-domain accumulation.
On an Intel Xeon Platinum 8468 for lengths $n=2^{10}$--$2^{18}$, the execution time is approximately 107--1315$\times$ that of FFTW's double-precision FFT, with NTTs accounting for approximately 80\% of the total time.

\end{abstract}

%%
%% The code below is generated by the tool at http://dl.acm.org/ccs.cfm.
%% Please copy and paste the code instead of the example below.
%%


%%
%% Keywords. The author(s) should pick words that accurately describe
%% the work being presented. Separate the keywords with commas.

\received{xx xxxx xxxx}
\received[revised]{xx xxxx xxxx}
\received[accepted]{xx xxxx xxxx}

%%
%% This command processes the author and affiliation and title
%% information and builds the first part of the formatted document.
\maketitle

\section{Introduction}

The fast Fourier transform (FFT)~\cite{Cooley1965}, an efficient algorithm for computing the discrete Fourier transform (DFT), is widely used in signal processing, scientific computing, and arbitrary-precision arithmetic.

Modern processors typically deliver much higher throughput for lower-precision arithmetic than for higher-precision arithmetic, and this gap continues to widen.
For example, from NVIDIA's H100 to B200 data center GPUs, the peak double-precision throughput increased only from 34 to 37~TFLOPS, whereas half-precision Tensor Core throughput increased from 990~TFLOPS to 2.2~PFLOPS and 8-bit Tensor Core throughput increased from 2.0~PFLOPS to 4.5~PFLOPS~\cite{NVIDIA-H100,NVIDIA-Blackwell-Arch}.
On recent consumer GPUs such as NVIDIA's GeForce RTX~5090, double-precision throughput is only 1/64 of single-precision throughput~\cite{NVIDIA-Blackwell}, since graphics workloads seldom require double precision.
Moreover, most processors provide no hardware support for floating-point arithmetic beyond double precision or for integer arithmetic beyond 64~bits.
High-precision formats such as quadruple precision (binary128)~\cite{IEEE754-2008}, Double-Double, and Quad-Double~\cite{Hida2001}, as well as arbitrary-precision libraries such as MPFR~\cite{Fousse2007} and GMP~\cite{GMP}, must therefore be realized in software at significantly reduced performance.

For matrix multiplication, the Ozaki scheme~\cite{Ozaki2012} offers a portable solution to this precision--performance gap: it computes matrix-matrix products at a target precision by splitting the inputs and delegating the core computation to an optimized lower-precision BLAS (Basic Linear Algebra Subprograms) routine.
Because achieving the target precision requires only an existing optimized BLAS library, no new kernel implementation or optimization is needed.
NVIDIA's next-generation Rubin architecture leverages the Ozaki scheme to achieve 200~TFLOPS of double-precision matrix multiplication on the R200 GPU, despite a native double-precision throughput of only 33~TFLOPS~\cite{NVIDIA-Rubin}.
Given that double-precision throughput is stagnating while low-precision throughput continues to increase, extending this approach beyond matrix multiplication to other fundamental operations is an important direction.

Research on computing FFTs at a target precision using lower-precision operations has focused primarily on achieving precision beyond double precision~\cite{VanDerHoeven2015,Sorna2018,mpFFT,Gong2022,Tolmachev2023,Kawakami2024}; however, all existing methods implement FFT kernels from scratch for each target precision.
No method analogous to the Ozaki scheme, namely one that computes FFTs at a target precision by reusing existing, optimized lower-precision FFTs, has been proposed.

In this paper, we propose such a method.
Our key observation is that the Bluestein FFT~\cite{Bluestein1970} reduces the DFT to a cyclic convolution, to which the Ozaki scheme can be applied.
The split component convolutions are then computed exactly using the number theoretic transform (NTT)~\cite{Pollard1971}, an FFT over a finite field, instead of a floating-point FFT, combined with the Chinese remainder theorem (CRT), thereby avoiding rounding errors in the convolution stage.

The contributions of this paper are as follows:
\begin{itemize}
  \item We propose a method for computing FFTs at a target precision using lower-precision FFTs by applying the Ozaki scheme to the cyclic convolution in the Bluestein FFT, with the NTT and CRT ensuring exact split component convolutions.
  \item We introduce an upper bound on the number of splits and an NTT-domain accumulation strategy, both of which reduce the number of NTT calls required.
  \item We implement a double-precision Bluestein FFT using 32-bit NTTs and evaluate its accuracy and computational cost, demonstrating that the Ozaki-scheme-based cyclic convolution effectively reduces the relative error of FFTs.
\end{itemize}

In this paper, we focus on establishing the algorithmic framework and evaluating its accuracy and computational cost.

The rest of this paper is organized as follows.
\Cref{sec:related-work} reviews related work.
\Cref{sec:preliminaries} introduces notation and background on FFTs, NTTs, CRT, and the Ozaki scheme.
\Cref{sec:proposed-method} presents the proposed method.
\Cref{sec:experiments} reports the experimental results.
\Cref{sec:discussion} discusses the conditions for competitive performance, extension to arbitrary-length FFTs, portability, and applicability to other precisions.
\Cref{sec:conclusion} concludes the paper.

\section{Related Work} \label{sec:related-work}

\subsection{Ozaki Scheme for Matrix Multiplication}

The Ozaki scheme~\cite{Ozaki2012} computes matrix-matrix products at a target precision using lower-precision matrix-matrix products.
It splits input matrices into components so that each split component product can be evaluated exactly by a standard BLAS routine, and then accumulates the results.
This approach has been applied in various settings to exploit hardware performance gaps.
Mukunoki et al.~\citeyearpar{Mukunoki2020} used half-precision Tensor Core matrix multiplication on NVIDIA GPUs to accelerate single- and double-precision matrix-matrix products, and Ootomo et al.~\citeyearpar{Ootomo2024} used 8-bit integer Tensor Core operations for the same purpose.
For precision beyond double, Mukunoki et al.~\citeyearpar{Mukunoki2021} used double-precision BLAS to accelerate quadruple-precision matrix multiplication, and Utsugiri and Kouya~\citeyearpar{Utsugiri2022} accelerated Triple-Double and Triple-Single (TS) precision matrix-matrix products using double- and single-precision BLAS routines, respectively.

\subsection{Computation of FFTs at Target Precision Using Lower-Precision Arithmetic}

Methods for computing FFTs at a target precision using lower-precision arithmetic have been developed primarily for precisions beyond double.
VkFFT~\cite{Tolmachev2023}, a cross-platform GPU FFT library, provides Double-Double-precision FFTs.
Gong et al.~\citeyearpar{Gong2022} implemented interval-arithmetic FFTs in Double-Double precision, and Kawakami and Takahashi~\citeyearpar{Kawakami2024} implemented Quad-Double-precision FFTs.
Hoeven and Lecerf~\citeyearpar{VanDerHoeven2015} implemented FFTs using fixed-point arithmetic at 4$\times$, 6$\times$, and 8$\times$ word length.
Frewer~\citeyearpar{mpFFT} developed mpFFT, an arbitrary-precision FFT library built on MPFR.

Sorna et al.~\citeyearpar{Sorna2018} applied a similar approach using half-precision Tensor Core operations on NVIDIA GPUs to compute FFTs more accurately than cuFFT's half-precision FFT.
All of these methods share a common approach: they first realize target-precision arithmetic using lower-precision operations and then implement dedicated FFT kernels from scratch.
Our method takes a different approach: by applying the Ozaki scheme to the cyclic convolution in the Bluestein FFT, it computes target-precision FFTs using NTTs (i.e., FFTs over a finite field) as the lower-precision FFT routine.

\section{Preliminaries} \label{sec:preliminaries}
\subsection{Notation}
We denote by $\mathbb{F}_{32}$ and $\mathbb{F}_{64}$ the sets of single-precision (binary32) and double-precision (binary64) floating-point numbers as defined by IEEE~754-2008~\cite{IEEE754-2008}, respectively; when the floating-point type need not be specified, we write $\mathbb{F}$.
We denote by $u$ the unit roundoff; for single precision, $u_{32} = 2^{-24}$, and for double precision, $u_{64} = 2^{-53}$.
We write $\mathbf{fl}_{32}(\cdot)$ and $\mathbf{fl}_{64}(\cdot)$ to denote the result of evaluating the argument in single- and double-precision arithmetic, respectively; when the floating-point type need not be specified, we write $\mathbf{fl}(\cdot)$.

For a vector $\bm{x}$, we write $x(j)$ for its $j$-th element.
We write $\Re(\cdot)$ and $\Im(\cdot)$ for the real and imaginary parts, respectively.
For two vectors $\bm{x}$ and $\bm{y}$, their elementwise product $\bm{z}$ with $z(j) = x(j)\,y(j)$ is denoted by $\bm{z} = \bm{x} \odot \bm{y}$.
We denote by $\bm{1}_n$ the $n$-dimensional all-ones vector.

The remainder of an integer $a$ modulo $m$ is computed to have minimum absolute value:
\begin{equation}
	a \bmod m = a - m \cdot \left\lfloor \frac{a}{m} + \frac{1}{2} \right\rfloor.
\end{equation}
That is, it satisfies
\begin{equation}
	-\frac{m}{2} \leq a \bmod m < \frac{m}{2}.
\end{equation}

\subsection{Fast Fourier Transform}
For a complex vector $\bm{x} \in \mathbb{C}^n$, the DFT is defined by
\begin{align}
	y(k)=\sum_{j=0}^{n-1} x(j)\,\omega_n^{jk}\quad (0\le k<n), \label{eq:dft}
\end{align}
where $y(k)\in\mathbb{C}$, $\omega_n=e^{-2\pi i/n}$, and $i=\sqrt{-1}$.
The inverse DFT is defined by
\begin{align}
	x(j)=\frac{1}{n}\sum_{k=0}^{n-1} y(k)\,\omega_n^{-jk}\quad (0\le j<n). \label{eq:idft}
\end{align}
We denote the DFT and inverse DFT of $\bm{x}$ by $\DFT(\bm{x})$ and $\DFT^{-1}(\bm{x})$, respectively.
By definition, $\DFT^{-1}$ is the inverse transform of the $\DFT$; that is, $\DFT^{-1}(\DFT(\bm{x}))=\bm{x}$.
Direct evaluation of \cref{eq:dft,eq:idft} requires $O\left(n^2\right)$ operations.

For composite lengths, the DFT can be factored into smaller DFTs, which leads to the well-known radix-based FFT algorithms.
By applying such factorizations recursively, algorithms such as the Cooley--Tukey FFT~\cite{Cooley1965} and the Stockham FFT~\cite{Cochran1967} compute an $n$-point DFT in $O(n\log n)$ time for lengths that admit suitable radix decompositions.
Hereafter, when the DFT and inverse DFT of $\bm{x}$ are computed by such an $O(n\log n)$ FFT routine and its inverse, we denote them by $\FFT(\bm{x})$ and $\FFT^{-1}(\bm{x})$, respectively; we write $\FFT_n(\bm{x})$ and $\FFT_n^{-1}(\bm{x})$ when the length must be stated explicitly.

However, for prime lengths or lengths with large prime factors, radix-based FFTs cannot be applied directly and efficiently, and thus alternative approaches are often employed.
A classical method that computes a DFT of arbitrary length in $O(n\log n)$ time is the Bluestein FFT~\cite{Bluestein1970} (also known as the chirp-$z$ transform).
The Bluestein FFT rewrites an $n$-point DFT as a convolution and evaluates it via an FFT-based cyclic convolution of a suitably chosen length $N$.

In the Bluestein FFT, \cref{eq:dft} is transformed as
\begin{align}
	y(k)
	= \sum_{j=0}^{n-1} x(j) \omega_n^{jk}
	= \omega_{2n}^{k^2} \sum_{j=0}^{n-1} \left(x(j) \omega_{2n}^{j^2}\right)\omega_{2n}^{-(k-j)^2}
	\quad (0\le k<n). \nonumber
\end{align}
Let $\bm{\omega}\in\mathbb{C}^n$ with $\omega(j)=\omega_{2n}^{j^2}$.
Then, $\bm{y}=\DFT(\bm{x})$ can be computed as follows:
\begin{enumerate}[label=Step~\arabic*., align=left, labelindent=\parindent, leftmargin=!]
	\item Compute the elementwise product: $\bm{x}'=\bm{x}\odot\bm{\omega}$.
	\item Compute the convolution:
	      \begin{equation}
		      y'(k)=\sum_{j=0}^{n-1} x'(j)\,\omega_{2n}^{-(k-j)^2}\quad (0\le k<n). \label{eq:bluestein_convolution}
	      \end{equation}
	\item Compute the elementwise product: $\bm{y}=\bm{y}'\odot\bm{\omega}$.
\end{enumerate}

To evaluate \cref{eq:bluestein_convolution} efficiently, its convolution sum is embedded into a cyclic convolution of length $N\ge 2n-1$.
For this embedding, we use two padding operators.
$\mathrm{ZeroPad}(\cdot,N)$ denotes standard zero-padding that extends a length-$n$ sequence to length $N$.
$\mathrm{ZeroPad}_{\pm}(\cdot,N)$ denotes zero-padding with index reversal (wrap-around).
Specifically, denoting zero-padded sequences by tilde, we define
\begin{align}
	\widetilde{\bm{x}'} &= \mathrm{ZeroPad}\!\left(\bm{x}',N\right)\in\mathbb{C}^N, \nonumber\\
	\widetilde{\bm{\omega}^{*}} &= \mathrm{ZeroPad}_{\pm}\!\left(\bm{\omega}^{*},N\right)\in\mathbb{C}^N, \nonumber
\end{align}
where $\bm{\omega}^{*}$ is the elementwise complex conjugate of $\bm{\omega}$, with $\omega^{*}(j)=\omega_{2n}^{-j^2}$.
In components,
\begin{align}
	\widetilde{x'}(j) &=
	\begin{cases}
		x'(j) & (0 \le j < n), \\
		0    & (n \le j < N),
	\end{cases} \nonumber\\
	\widetilde{\omega^{*}}(j) &=
	\begin{cases}
		\omega^{*}(j)       & (0 \le j < n), \\
		0                  & (n \le j < N - n + 1), \\
		\omega^{*}(N - j) & (N - n + 1 \le j < N).
	\end{cases} \nonumber
\end{align}
Then, the cyclic convolution can be computed by FFTs of length $N$ as
\begin{align}
	\widetilde{\bm{y}'}=\FFT_N^{-1}\!\left(\FFT_N\!\left(\widetilde{\bm{x}'}\right)\odot \FFT_N\!\left(\widetilde{\bm{\omega}^{*}}\right)\right),
\end{align}
and the desired values $\bm{y}'$ are obtained as the first $n$ entries of $\widetilde{\bm{y}'}$.
By choosing $N$ to be a length for which an $O(N\log N)$ FFT is available, the cyclic convolution in Step~2 can be computed in $O(N\log N)$ time.
Including the $O(n)$ elementwise products in Steps~1 and~3, the overall complexity of the Bluestein FFT is $O(N\log N + n)=O(n\log n)$.
The algorithm of the Bluestein FFT is summarized in \cref{alg:bluestein_fft}.
\begin{algorithm}[t]
	\caption{Bluestein FFT}
	\label{alg:bluestein_fft}
	\begin{algorithmic}[1]
		\Require $\bm{x}\in \mathbb{C}^{n}$, $\bm{\omega}\in\mathbb{C}^{n}$ with $\omega(j)=\omega_{2n}^{j^2}$, $\FFT_{N}$, $\FFT_{N}^{-1}$ with $N \ge 2n-1$
		\Ensure $\bm{y}=\DFT(\bm{x})\in \mathbb{C}^{n}$
		\State $\bm{x}' \gets \bm{x} \odot \bm{\omega}$
		\State $\widetilde{\bm{x}'} \gets \mathrm{ZeroPad}\!\left(\bm{x}',N\right)$
		\State $\widetilde{\bm{\omega}^{*}} \gets \mathrm{ZeroPad}_{\pm}\!\left(\bm{\omega}^{*},N\right)$
		\State $\widetilde{\bm{y}'} \gets \FFT_{N}^{-1}\!\left(\FFT_{N}\!\left(\widetilde{\bm{x}'}\right) \odot \FFT_{N}\!\left(\widetilde{\bm{\omega}^{*}}\right)\right)$
		\State $\bm{y}' \gets$ the first $n$ entries of $\widetilde{\bm{y}'}$
		\State $\bm{y} \gets \bm{y}' \odot \bm{\omega}$
	\end{algorithmic}
\end{algorithm}

When $n$ is a power of two, we can typically compute the DFT using the Cooley--Tukey FFT or the Stockham FFT, and thus the Bluestein FFT is usually not employed.
Nevertheless, if the Bluestein FFT is applied to such a length, the convolution in \cref{eq:bluestein_convolution} becomes a cyclic convolution; therefore, zero-padding is unnecessary and the FFT length $N$ can be kept equal to $n$.
In this case, the Bluestein FFT evaluates the DFT via a cyclic convolution using two length-$n$ forward FFTs and one length-$n$ inverse FFT, in addition to elementwise products.
Although the asymptotic complexity remains $O(n\log n)$, the constant factor is approximately three times as large as that of a radix-based FFT such as the Cooley--Tukey or Stockham algorithm.

\subsection{Number Theoretic Transform and Chinese Remainder Theorem}
The NTT~\cite{Pollard1971} is a DFT defined over the finite field
$\mathbb{Z}/p\mathbb{Z}=\mathbb{Z}_p$ for a prime modulus $p$.
For a length-$n$ vector $\bm{x}\in\mathbb{Z}_p^n$, the NTT is defined by
\begin{align}
  y(k)=\sum_{j=0}^{n-1} x(j)\,\omega_n^{jk}\bmod p \quad (0\le k<n), \label{eq:ntt}
\end{align}
where $\omega_n\in\mathbb{Z}_p$ is a primitive $n$-th root of unity; that is,
$\omega_n^n\equiv 1\ (\bmod\,p)$ and $\omega_n^j\not\equiv 1\ (\bmod\,p)$ for $0<j<n$.
The inverse NTT is defined by
\begin{align}
  x(j) = n^{-1}\sum_{k=0}^{n-1} y(k)\,\omega_n^{-jk}\bmod p \quad (0\le j<n), \label{eq:intt}
\end{align}
where $n^{-1}$ and $\omega_n^{-jk}$ denote the multiplicative inverses of $n$ and $\omega_n^{jk}$ in $\mathbb{Z}_p$, respectively.
We denote the NTT and the inverse NTT modulo $p$ by $\NTT_p(\bm{x})$ and $\NTT_p^{-1}(\bm{x})$, respectively.
When the modulus is clear from the context, we omit the subscript $p$ and write $\NTT(\bm{x})$ and $\NTT^{-1}(\bm{x})$.
By definition, the inverse NTT is the inverse transform of the NTT; that is, $\NTT_p^{-1}(\NTT_p(\bm{x}))=\bm{x}$.

Direct evaluation of \eqref{eq:ntt} and \eqref{eq:intt} requires $O\left(n^2\right)$ operations.
As in the complex case, when $n$ admits suitable factorizations, Cooley--Tukey- or Stockham-type decompositions
can be used to compute the NTT in $O(n\log n)$ time.
Hereafter, $\NTT(\cdot)$ and $\NTT^{-1}(\cdot)$ denote the NTT and the inverse NTT computed by an $O(n\log n)$ algorithm, respectively.

For $\bm{x},\bm{y}\in\mathbb{Z}_p^n$, the cyclic convolution $\bm{z}=\bm{x}\circledast \bm{y}\bmod p$
can be computed via the convolution theorem as
\begin{align}
  \bm{z}=\NTT_p^{-1}\left(\NTT_p(\bm{x})\odot \NTT_p(\bm{y})\right).
\end{align}
If all entries of $\bm{x}$, $\bm{y}$, and $\bm{x}\circledast\bm{y}$ lie in $[-p/2,p/2)$,
then the integer cyclic convolution $\bm{x}\circledast\bm{y}$ can be computed exactly using the NTT.

Let $m_0,\ldots,m_{d-1}$ be pairwise coprime integers and let $M=\prod_{j=0}^{d-1} m_j$.
The CRT states that, given residues $a\equiv a_j \ (\bmod\,m_j)$ for $0\le j< d$,
the solution $a$ is uniquely determined modulo $M$.
One explicit reconstruction is
\begin{equation}
  a \equiv \sum_{j=0}^{d-1} a_j\,M_j\,M_j^{-1} \ (\bmod\,M),
\end{equation}
where $M_j = M/m_j$ and $M_j^{-1}$ denotes the multiplicative inverse of $M_j$ modulo $m_j$.

The CRT can be used to reconstruct integer cyclic convolutions computed by NTTs with different moduli.
Let $p_0,\ldots,p_{d-1}$ be distinct primes and let $P=\prod_{j=0}^{d-1} p_j$.
To compute the cyclic convolution of integer vectors $\bm{x}$ and $\bm{y}$ modulo $P$,
we first compute the results modulo each $p_j$:
\begin{align}
  \bm{z}_j
  =\NTT_{p_j}^{-1}\!\left(
    \NTT_{p_j}\!\left(\bm{x}\bmod p_j\right)\odot
    \NTT_{p_j}\!\left(\bm{y}\bmod p_j\right)
  \right)\quad (0\le j<d).
\end{align}
Then, $\bm{z}=\bm{x}\circledast \bm{y}\bmod P$ is obtained by CRT reconstruction:
\begin{equation}
  \bm{z}
  \equiv
  \sum_{j=0}^{d-1} \bm{z}_j\,P_j\,P_j^{-1} \ (\bmod\,P),
  \label{eq:crt_ntt}
\end{equation}
where $P_j = P/p_j$ and $P_j^{-1}$ denotes the multiplicative inverse of $P_j$ modulo $p_j$.

Instead of evaluating \cref{eq:crt_ntt} directly, we may reconstruct $\bm{z}$ incrementally via Garner's algorithm~\cite{Garner1959}.
Let $P_{j}^{\prime}=\prod_{k=0}^{j} p_k$ and set $\bm{z}'_0=\bm{z}_0$.
For $1\le j<d$, define
\begin{align}
  \bm{z}'_{j}
  =
  \bm{z}'_{j-1}
  + \left\{\left(\bm{z}_j-\bm{z}'_{j-1}\right)\left(P_{j-1}^{\prime}\right)^{-1}\bmod p_j\right\}P_{j-1}^{\prime},
\end{align}
where $\left(P_{j-1}^{\prime}\right)^{-1}$ denotes the multiplicative inverse of $P_{j-1}^{\prime}$ modulo $p_j$.
Then, $\bm{z}'_{d-1}\equiv \bm{z} \ (\bmod\,P)$ holds.

\subsection{Ozaki Scheme}
The Ozaki scheme is an error-free transformation for matrix multiplication.
In this paper, for simplicity of exposition, we describe the Ozaki scheme for dot products; the same idea extends to matrix-vector and matrix-matrix products.
The Ozaki scheme evaluates a dot product in the following three steps:
\begin{enumerate}[label=Step~\arabic*., align=left, labelindent=\parindent, leftmargin=!]
	\item Split the input vectors:
	\begin{align}
		\bm{x} \rightarrow \bm{x}^{(0)} + \cdots + \bm{x}^{(k_x - 1)},
		\quad
		\bm{y} \rightarrow \bm{y}^{(0)} + \cdots + \bm{y}^{(k_y - 1)}.
	\end{align}
	\item Compute the dot products of the splits:
	\begin{align}
		z^{(s, t)} = \bm{x}^{(s)} \cdot \bm{y}^{(t)}
		\quad (0 \le s < k_x,\ 0 \le t < k_y).
	\end{align}
	\item Accumulate the results:
	\begin{align}
		z = \sum_{s=0}^{k_x - 1} \sum_{t=0}^{k_y - 1} z^{(s, t)}.
	\end{align}
\end{enumerate}
The dot-product algorithm based on the Ozaki scheme is summarized in \cref{alg:ozaki-scheme}.
\begin{algorithm}[t]
	\caption{Dot product using Ozaki scheme}
	\label{alg:ozaki-scheme}
	\begin{algorithmic}[1]
		\Require $\bm{x}, \bm{y} \in \mathbb{F}^{n}$
		\Ensure $z \simeq \bm{x} \cdot \bm{y} \in \mathbb{F}$
		\State $\bm{x}^{(0)}, \ldots, \bm{x}^{(k_x - 1)} \in \mathbb{F}^{n} \gets \mathrm{Split}~(\bm{x})$
		\State $\bm{y}^{(0)}, \ldots, \bm{y}^{(k_y - 1)} \in \mathbb{F}^{n} \gets \mathrm{Split}~(\bm{y})$
		\For {$s= 0, \ldots, k_x - 1$}
			\For {$t = 0, \ldots, k_y - 1$}
				\State $z^{(s, t)} \gets \bm{x}^{(s)} \cdot \bm{y}^{(t)}$
			\EndFor
		\EndFor
		\State $z \gets \sum_{s=0}^{k_x - 1} \sum_{t=0}^{k_y - 1} z^{(s, t)}$
	\end{algorithmic}
\end{algorithm}

In Step~1, \cref{alg:ozaki-scheme-split} splits floating-point vectors $\bm{x},\bm{y}\in\mathbb{F}^n$ into sums of $k_x$ and $k_y$ fixed-point vectors with $\alpha$ bits, where
\begin{equation}
	\alpha = \left\lfloor \frac{\log_2\!\left(u^{-1}\right) - \log_2\left(n\right)}{2} \right\rfloor.
	\label{eq:ozaki-scheme-alpha}
\end{equation}
More precisely, with the scaling factors $\sigma^{\left(k\right)}$ defined in \cref{alg:ozaki-scheme-split}, the split components of $\bm{x}$ satisfy the following conditions:
\begin{equation}
	\left\|\bm{x}^{\left(k\right)}\right\|_\infty \leq 2^{\alpha} u \sigma^{\left(k\right)}, \quad \bm{x}^{\left(k\right)} \in u \sigma^{\left(k\right)} \mathbb{Z}^n \quad \left(0 \leq k < k_x\right).
	\label{eq:ozaki-scheme-split-condition}
\end{equation}
The same conditions apply to the split components of $\bm{y}$.
With this choice of $\alpha$, the split is constructed such that no rounding error occurs in the dot products in Step~2; that is,
\begin{equation}
	\bm{x}^{\left(s\right)} \cdot \bm{y}^{\left(t\right)} = \mathbf{fl}\!\left(\bm{x}^{\left(s\right)} \cdot \bm{y}^{\left(t\right)}\right)
\end{equation}
for all $\left(s,t\right)$.
\begin{algorithm}[t]
	\caption{Splitting procedure in Ozaki scheme}
	\label{alg:ozaki-scheme-split}
	\begin{algorithmic}[1]
		\Require $\bm{x} \in \mathbb{F}^{n}$
		\Ensure $\bm{x}^{(0)}, \ldots, \bm{x}^{(k - 1)} \in \mathbb{F}^{n}$ such that $\sum_{j = 0}^{k - 1} \bm{x}^{(j)} = \bm{x}$
		\Function{$\mathrm{Split}$~}{$\bm{x}$}
			\State $\rho \gets \left\lceil \left(\log_2 \left(u^{-1}\right) + \log_2 (n)\right) / 2 \right\rceil$
			\State $\bm{r}^{(0)} \gets \bm{x}$
			\State $\mu^{(0)} \gets \left\|\bm{r}^{(0)}\right\|_\infty$
			\State $k \gets 0$
			\While {$\mu^{(k)} > 0$}
				\State $\sigma^{(k)} \gets 2^{\left\lceil \log_2\left(\mu^{(k)}\right) \right\rceil + \rho}$
				\State $\bm{x}^{(k)} \gets \mathbf{fl} \left(\left(\bm{r}^{(k)} + \sigma^{(k)} \bm{1}_n \right) - \sigma^{(k)} \bm{1}_n\right)$
				\State $\bm{r}^{(k + 1)} \gets \bm{r}^{(k)} - \bm{x}^{(k)}$
				\State $\mu^{(k + 1)} \gets \left\|\bm{r}^{(k + 1)}\right\|_\infty$
				\State $k \gets k + 1$
			\EndWhile
			\State \Return $\bm{x}^{(0)}, \ldots, \bm{x}^{(k - 1)}$
		\EndFunction
	\end{algorithmic}
\end{algorithm}

The number of splits depends on the distribution of vector entries and on $\alpha$.
With $|\bm{x}|$ denoting the elementwise absolute value of $\bm{x}$, an upper bound on the number of splits $k_x$ is given by
\begin{align}
	k_x \le \left\lceil
	\frac{\log_2\!\left(\max |\bm{x}|\right) - \log_2\!\left(\min |\bm{x}|\right) + \log_2\!\left(u^{-1}\right) - 1}{\alpha - 1}
	\right\rceil.
	\label{eq:k-upper-limit}
\end{align}
Thus, $k_x$ increases as the exponent range of the entries of $\bm{x}$ becomes larger.
Moreover, $k_x$ tends to increase when $\alpha$ is smaller; that is, when $n$ is larger or when the mantissa length $\log_2(u^{-1})$ is smaller.

When $k_x = k_y = K$, the computational costs of Steps~1 to 3 for a dot product are $O(Kn)$, $O(K^2n)$, and $O(K^2)$, respectively.
For matrix multiplication, the corresponding costs are $O(Kn^2)$, $O(K^2n^3)$, and $O(K^2n^2)$, respectively.
Therefore, Step~2 is the dominant step, and the overall complexity is $O(K^2n)$ for dot products and $O(K^2n^3)$ for matrix multiplication.
Consequently, the computational cost is governed by the number of splits $K$, which depends on the split width $\alpha$ and the dynamic range of the input values, as discussed above.

\section{Proposed Method} \label{sec:proposed-method}

In this section, we present our method for computing FFTs at a target precision using lower-precision FFTs.
Throughout the discussion, we assume that the DFT size $n$ is a power of two.

\subsection{Motivation: Rationale for Using Bluestein FFT}
Prior work on computing FFTs with higher precision than that directly supported by hardware typically follows a two-step approach: (1) emulate high-precision arithmetic using multiple lower-precision values and operations, and (2) implement FFT algorithms using these emulated operations.
For instance, Double-Double and Quad-Double arithmetic~\cite{Hida2001} use pairs or quadruples of double-precision numbers to achieve mantissas of 106 or 212 bits, respectively. FFTs have been implemented using these data types.

To compute double-precision FFTs using lower-precision arithmetic, following this paradigm, we might employ TS arithmetic~\cite{Fabiano2019}, which represents a floating-point number using three single-precision values.
TS arithmetic provides a mantissa of $24 \times 3 = 72$ bits, exceeding the 53-bit mantissa of double precision, though its exponent range remains limited to that of single precision.

A TS number represents the true value $x$ as
\begin{equation}
	x = x_{0} + x_{1} + x_{2},
\end{equation}
where $x_{0}, x_{1}, x_{2} \in \mathbb{F}_{32}$ and satisfy
\begin{equation}
	x_{s-1} \neq 0 \implies \left|x_{s}\right| \leq \frac{1}{2}\,\mathrm{ulp}\left(x_{s-1}\right), \quad s = 1, 2.
\end{equation}
Here, $\mathrm{ulp}(x)$ denotes the unit in the last place of $x$.
We write $x_{0:2}$ as shorthand for the triple $(x_0, x_1, x_2)$ and refer to it as a TS number.

Now, consider computing a double-precision DFT using TS arithmetic.
Let $\bm{x} \in \mathbb{F}_{64}^{n} + i \mathbb{F}_{64}^{n}$ be a double-precision complex vector and $\bm{\omega} \in \mathbb{F}_{64}^{n} + i \mathbb{F}_{64}^{n}$ be the twiddle factor vector with $\omega(j) = \mathbf{fl}_{64}\left(\omega_n^{j}\right)$.
We can split each of them into three single-precision complex vectors
$\bm{x}_{s},\, \bm{\omega}_{t} \in \mathbb{F}_{32}^{n} + i\mathbb{F}_{32}^{n}$ $(s, t = 0, 1, 2)$.
The DFT of $\bm{x}$ can then be expressed as
\begin{align}
	y(k) & = \sum_{j = 0}^{n - 1} x(j) \omega_n^{jk} \nonumber\\
	& \simeq \sum_{j = 0}^{n - 1} \left(\sum_{s=0}^{2} x_{s}(j)\right) \left(\sum_{t=0}^{2} \omega_{t}(jk \bmod n)\right) \quad (0 \leq k < n). \label{eq:ts-fft}
\end{align}
We could implement the Cooley--Tukey or Stockham FFT using TS arithmetic to evaluate \cref{eq:ts-fft} in $O(n\log n)$ time.
However, expanding \cref{eq:ts-fft} reveals:
\begin{align}
	  & \sum_{j = 0}^{n - 1} \left(\sum_{s=0}^{2} x_{s}(j)\right) \left(\sum_{t=0}^{2} \omega_{t}(jk \bmod n)\right) \quad (0 \leq k < n) \nonumber \\
	= & \sum_{s = 0}^{2} \sum_{t = 0}^{2} \sum_{j = 0}^{n - 1} x_{s}(j) \omega_{t}(jk \bmod n) \quad (0 \leq k < n). \nonumber
\end{align}
Each term $\sum_{j = 0}^{n - 1} x_{s}(j) \omega_{t}(jk \bmod n)$ lacks a DFT structure, so we cannot apply single-precision FFT routines to compute all terms simultaneously.
Our goal is to compute a target-precision FFT using only lower-precision FFT routines, which benefit from highly optimized hardware implementations.
To achieve this goal, we turn to the Bluestein FFT algorithm, which reformulates the DFT as a cyclic convolution.
We consider computing a double-precision DFT using TS arithmetic with the Bluestein FFT approach.
The algorithm proceeds in the following steps:
\begin{enumerate}[label=Step~\arabic*., align=left, labelindent=\parindent, leftmargin=!]
	\item Convert the double-precision complex vectors $\bm{x}$, $\bm{\omega}$, and $\bm{\omega}^{*}$ to TS complex vectors.
	\item Compute the elementwise product $\bm{x}' \gets \bm{x} \odot \bm{\omega}$ with TS arithmetic.
	\item Compute the cyclic convolution $\bm{y}' \gets \bm{x}' \circledast \bm{\omega}^{*}$ with TS arithmetic.
	\item Compute the elementwise product $\bm{y} \gets \bm{y}' \odot \bm{\omega}$ with TS arithmetic.
	\item Convert the result $\bm{y}$ from a TS complex vector back to a double-precision complex vector.
\end{enumerate}
The above procedure is summarized in \cref{alg:ts-bluestein-fft}.
\begin{algorithm}[t]
	\caption{Double-precision Bluestein FFT using TS arithmetic. Here, $\mathrm{TS}(\cdot)$ converts a double-precision value to its TS representation, and $\mathrm{TSAdd}$, $\mathrm{TSSub}$, and $\mathrm{TSMul}$ denote TS addition, subtraction, and multiplication, respectively.}
	\label{alg:ts-bluestein-fft}
	\begin{algorithmic}[1]
		\Require $\bm{x}, \bm{\omega}, \bm{\omega}^* \in \mathbb{F}^{n}_{64} + i\mathbb{F}^{n}_{64}$
		\Ensure $\bm{y} \in \mathbb{F}^{n}_{64} + i\mathbb{F}^{n}_{64}$ such that $\bm{y} \simeq \DFT(\bm{x})$
		
		\Statex \textbf{1. Convert inputs to TS.}
		\State $\bm{x}_{0:2} \gets \mathrm{TS}\left(\bm{x}\right)$
		\State $\bm{\omega}_{0:2} \gets \mathrm{TS}\left(\bm{\omega}\right)$
		\State $\bm{\omega}_{0:2}^{*} \gets \mathrm{TS}\left(\bm{\omega}^{*}\right)$

		\Statex \textbf{2. Elementwise product in TS: $\bm{x}' \gets \bm{x} \odot \bm{\omega}$.}
		\State $\Re\left(\bm{x}^{\prime}_{0:2}\right) \gets \mathrm{TSSub}\left(\mathrm{TSMul}\left(\Re\left(\bm{x}_{0:2}\right), \Re\left(\bm{\omega}_{0:2}\right)\right), \mathrm{TSMul}\left(\Im\left(\bm{x}_{0:2}\right), \Im\left(\bm{\omega}_{0:2}\right)\right)\right)$
		\State $\Im\left(\bm{x}^{\prime}_{0:2}\right) \gets \mathrm{TSAdd}\left(\mathrm{TSMul}\left(\Re\left(\bm{x}_{0:2}\right), \Im\left(\bm{\omega}_{0:2}\right)\right), \mathrm{TSMul}\left(\Im\left(\bm{x}_{0:2}\right), \Re\left(\bm{\omega}_{0:2}\right)\right)\right)$

		\Statex \textbf{3. Cyclic convolution in TS: $\bm{y}' \gets \bm{x}' \circledast \bm{\omega}^{*}$.}
		\State $\bm{y}^{\prime}_{0:2} \gets \bm{x}^{\prime}_{0:2} \circledast \bm{\omega}^{*}_{0:2}$

		\Statex \textbf{4. Elementwise product in TS: $\bm{y} \gets \bm{y}' \odot \bm{\omega}$.}
		\State $\Re\left(\bm{y}_{0:2}\right) \gets \mathrm{TSSub}\left(\mathrm{TSMul}\left(\Re\left(\bm{y}_{0:2}^{\prime}\right), \Re\left(\bm{\omega}_{0:2}\right)\right), \mathrm{TSMul}\left(\Im\left(\bm{y}_{0:2}^{\prime}\right), \Im\left(\bm{\omega}_{0:2}\right)\right)\right)$
		\State $\Im\left(\bm{y}_{0:2}\right) \gets \mathrm{TSAdd}\left(\mathrm{TSMul}\left(\Re\left(\bm{y}_{0:2}^{\prime}\right), \Im\left(\bm{\omega}_{0:2}\right)\right), \mathrm{TSMul}\left(\Im\left(\bm{y}_{0:2}^{\prime}\right), \Re\left(\bm{\omega}_{0:2}\right)\right)\right)$

		\Statex \textbf{5. Convert TS result to FP64.}
		\State $\bm{y} \gets \fl_{64}\left(\bm{y}_{0} + \bm{y}_{1} + \bm{y}_{2}\right)$
	\end{algorithmic}
\end{algorithm}

Now, consider the cyclic convolution in Step~3 between two TS vectors
$\bm{x}_{s},\, \bm{y}_{t} \in \mathbb{F}_{32}^{n}$ $(s, t = 0, 1, 2)$.
This takes the form:
\begin{align}
	z(k) & = \sum_{j = 0}^{n - 1} x(j)\, y((k - j) \bmod n)\nonumber\\
	& \simeq \sum_{j = 0}^{n - 1}  \left(\sum_{s=0}^{2} x_{s}(j)\right)\left(\sum_{t=0}^{2} y_{t}((k - j) \bmod n)\right) \quad (0 \leq k < n). \label{eq:ts-conv}
\end{align}
We could compute \cref{eq:ts-conv} using FFT-based convolution with the Cooley--Tukey or Stockham FFT in TS arithmetic in $O(n \log n)$ time.
However, expanding \cref{eq:ts-conv} yields:
\begin{align}
	  & \sum_{j = 0}^{n - 1}  \left(\sum_{s=0}^{2} x_{s}(j)\right)\left(\sum_{t=0}^{2} y_{t}((k - j) \bmod n)\right) \nonumber \\
	= & \sum_{s = 0}^{2} \sum_{t = 0}^{2} \sum_{j = 0}^{n - 1} x_{s}(j)\, y_{t}((k - j) \bmod n) \quad (0 \leq k < n). \nonumber
\end{align}
Crucially, $\sum_{j = 0}^{n - 1} x_{s}(j)\, y_{t}((k - j) \bmod n)$ is a cyclic convolution between single-precision vectors $\bm{x}_{s}$ and $\bm{y}_{t}$.
Thus, we can compute it in $O(n \log n)$ time using single-precision FFT routines.

However, a naive implementation using single-precision FFT yields only single-precision accuracy for each term, so even when all terms are summed to form the TS-precision result, errors are introduced.
This is where the Ozaki scheme becomes essential: if we can apply the Ozaki scheme to cyclic convolution, we can split the TS vectors into multiple single-precision vectors such that each split component cyclic convolution is computed exactly (without rounding error) using single-precision FFTs.
This is the key insight underlying the proposed method.

\subsection{Application of Ozaki Scheme to Cyclic Convolution}
For floating-point vectors $\bm{x}, \bm{y} \in \mathbb{F}^n$, the cyclic convolution
$\bm{z} = \bm{x} \circledast \bm{y}$ can be expressed as a matrix-vector product
$\bm{z} = Y\bm{x}$ using the circulant matrix generated by $\bm{y}$:
\begin{equation}
	Y = \left(y_{jk}\right) = \begin{pmatrix}
		y(0)      & y(n - 1) & \cdots & y(1)    \\
		y(1)       & y(0)       & \cdots & y(2)    \\
		\vdots    & \vdots    & \ddots & \vdots \\
		y(n - 1) & y(n - 2) & \cdots & y(0)
	\end{pmatrix} \in \mathbb{F}^{n \times n}.
\end{equation}
Since $y_{jk} = y\left((j - k) \bmod n\right)$, each row of $Y$ is a cyclic shift of the entries of $\bm{y}$.
Hence, the Ozaki scheme can be applied to cyclic convolution by treating the convolution as a matrix-vector product.

In \cref{alg:ozaki-scheme-split}, the quantities $\mu^{(k)}$ and $\sigma^{(k)}$ are independent of the ordering of the entries.
Therefore, if applying \cref{alg:ozaki-scheme-split} to $\bm{y}$ yields a splitting
$\bm{y}=\sum_{t=0}^{k_y-1}\bm{y}^{(t)}$, then we can define
$Y=\sum_{t=0}^{k_y-1}Y^{(t)}$ with entries $y^{(t)}_{jk}=y^{(t)}\left((j-k) \bmod n\right)$, where each $Y^{(t)}$ is a circulant matrix.
Hence, if $\bm{x}$ is split by \cref{alg:ozaki-scheme-split} as
$\bm{x} = \sum_{s = 0}^{k_x - 1} \bm{x}^{(s)}$, then the cyclic convolution
$\bm{z} = \bm{x} \circledast \bm{y}$ can be computed as
\begin{equation}
	\bm{z}
  = \sum_{s = 0}^{k_x - 1} \sum_{t = 0}^{k_y - 1} Y^{(t)} \bm{x}^{(s)}
  = \sum_{s = 0}^{k_x - 1} \sum_{t = 0}^{k_y - 1} \bm{x}^{(s)} \circledast \bm{y}^{(t)}.
  \label{eq:ozaki-scheme-conv}
\end{equation}
A direct evaluation of $Y^{(t)}\bm{x}^{(s)}$ requires $O(n^2)$ time, whereas each cyclic convolution term
$\bm{x}^{(s)} \circledast \bm{y}^{(t)}$ can be computed in $O(n \log n)$ time using FFT-based convolution.

When FFT-based convolution is used, each term in \cref{eq:ozaki-scheme-conv} is evaluated as
\begin{equation}
  \bm{x}^{(s)} \circledast \bm{y}^{(t)}
  = \FFT^{-1}\!\left(\FFT\!\left(\bm{x}^{(s)}\right)\odot \FFT\!\left(\bm{y}^{(t)}\right)\right).
\end{equation}
It is a real-valued cyclic convolution; hence, when computed with floating-point FFTs, rounding errors will occur.

In contrast, for an integer convolution computed with FFTs, if the exact result is bounded by $u^{-1}$ and the numerical error is strictly less than $1/2$, then the exact integer result can be recovered by rounding to the nearest integer.
By \cref{eq:ozaki-scheme-split-condition}, each split component can be scaled by $\left(u\sigma^{(k)}\right)^{-1}$ to an integer vector with entries bounded by $2^{\alpha}$.
We therefore adopt this integer approach.
Let $c_x^{(s)}=\left(u\sigma^{(s)}\right)^{-1}$ and $c_y^{(t)}=\left(u\sigma^{(t)}\right)^{-1}$ be the scaling factors obtained from splitting $\bm{x}$ and $\bm{y}$, respectively.
Then, $\bm{x}^{(s)} \circledast \bm{y}^{(t)}$ can be computed as
\begin{equation}
	\bm{x}^{(s)} \circledast \bm{y}^{(t)}
  = \left.
  \left\lfloor
  \FFT^{-1}\!\left(
    \FFT\!\left(c_x^{(s)}\bm{x}^{(s)}\right)\odot
    \FFT\!\left(c_y^{(t)}\bm{y}^{(t)}\right)
  \right)
  \right\rceil
  \middle/
  \left(c_x^{(s)} c_y^{(t)}\right)
  \right.,
\end{equation}
where $\lfloor\cdot\rceil$ denotes rounding to the nearest integer.

However, when an integer cyclic convolution is computed by floating-point FFTs, the error is not necessarily below $1/2$.
An upper bound on the maximum error of FFT-based convolution is given by \cite{Percival2003}
\begin{equation}
	\left\|\bm{z}' - \bm{z}\right\|_{\infty}
  \leq
  \left\|c_x^{(s)}\bm{x}^{(s)}\right\|_2 \cdot \left\|c_y^{(t)}\bm{y}^{(t)}\right\|_2
  \cdot
  \left\{
    \left(1 + u\right)^{3n}
    \left(1 + u\sqrt{5}\right)^{3n + 1}
    \left(1 + \|\bm{\omega}' - \bm{\omega}\|_\infty \right)^{3n}
    - 1
  \right\},
\end{equation}
where $\bm{z}'$ is the result computed by FFT-based cyclic convolution and $\bm{z}$ is the exact value.
Here, $\bm{\omega}'$ denotes the trigonometric table used in the FFT and $\bm{\omega}$ denotes the exact trigonometric values; thus, $\left\|\bm{\omega}'-\bm{\omega}\right\|_\infty$ is the maximum table error.
Assuming correctly rounded table entries, we have $\left\|\bm{\omega}'-\bm{\omega}\right\|_\infty \leq u / \sqrt{2}$.

If $\left\|c_x^{(s)}\bm{x}^{(s)}\right\|_\infty,\ \left\|c_y^{(t)}\bm{y}^{(t)}\right\|_\infty \le 2^{\alpha}$, then
$\left\|c_x^{(s)}\bm{x}^{(s)}\right\|_2 \cdot \left\|c_y^{(t)}\bm{y}^{(t)}\right\|_2 \le n2^{2\alpha}$.
To ensure correct rounding to the exact result, we require $\|\bm{z}'-\bm{z}\|_\infty < 1/2$; that is,
\begin{equation}
	\|\bm{z}' - \bm{z}\|_{\infty}
  \leq n2^{2\alpha}
  \cdot
  \left\{
    \left(1 + u\right)^{3n}
    \left(1 + u\sqrt{5}\right)^{3n + 1}
    \left(1 + \frac{u}{\sqrt{2}} \right)^{3n}
    - 1
  \right\}
  < \frac{1}{2}.
\end{equation}
Solving this inequality for $\alpha$ yields
\begin{align}
	\alpha
  < -\frac{1}{2}
  \left[
    1 + \log_2(n)
    + \log_2\left\{
      \left(1 + u\right)^{3n}
      \left(1 + u\sqrt{5}\right)^{3n + 1}
      \left(1 + \frac{u}{\sqrt{2}} \right)^{3n}
      - 1
    \right\}
  \right].
\end{align}
Therefore, it suffices to choose
\begin{equation}
	\alpha =
  \left\lfloor
  -\frac{1}{2}
  \left[
    1 + \log_2(n)
    + \log_2\left\{
      \left(1 + u\right)^{3n}
      \left(1 + u\sqrt{5}\right)^{3n + 1}
      \left(1 + \frac{u}{\sqrt{2}} \right)^{3n}
      - 1
    \right\}
  \right]
  \right\rfloor.
  \label{eq:ozaki-scheme-conv-fft-alpha}
\end{equation}
This condition is more restrictive than \cref{eq:ozaki-scheme-alpha}, which determines $\alpha$ in the original Ozaki scheme.

In contrast, using the NTT allows us to compute integer cyclic convolutions exactly.
With an NTT modulo $p$, $\bm{x}^{(s)} \circledast \bm{y}^{(t)}$ can be evaluated as
\begin{equation}
	\bm{x}^{(s)} \circledast \bm{y}^{(t)}
  = \left.
  \NTT^{-1}_{p}\!\left(
    \NTT_{p}\!\left(c_x^{(s)}\bm{x}^{(s)}\right)\odot
    \NTT_{p}\!\left(c_y^{(t)}\bm{y}^{(t)}\right)
  \right)
  \middle/
  \left(c_x^{(s)} c_y^{(t)}\right)
  \right..
\end{equation}
Since $\left\|c_x^{(s)}\bm{x}^{(s)} \circledast c_y^{(t)}\bm{y}^{(t)}\right\|_\infty \le n\,\left\|c_x^{(s)}\bm{x}^{(s)}\right\|_\infty \cdot \left\|c_y^{(t)}\bm{y}^{(t)}\right\|_\infty$, a sufficient condition for exact recovery is
\begin{equation}
  n\,\left\|c_x^{(s)}\bm{x}^{(s)}\right\|_\infty \cdot
    \left\|c_y^{(t)}\bm{y}^{(t)}\right\|_\infty
  < \frac{p}{2}.
\end{equation}
Furthermore, if
$\left\|c_x^{(s)}\bm{x}^{(s)}\right\|_\infty,
 \left\|c_y^{(t)}\bm{y}^{(t)}\right\|_\infty \le 2^\alpha$,
then
\begin{equation}
  n\,2^{2\alpha} < \frac{p}{2},
\end{equation}
which implies
\begin{equation}
	\alpha < \frac{\log_2(p/2) - \log_2(n)}{2}.
\end{equation}
Therefore, it suffices to choose
\begin{equation}
	\alpha = \left\lfloor \frac{\log_2(p/2) - \log_2(n)}{2} \right\rfloor.
  \label{eq:ozaki-scheme-conv-ntt-alpha}
\end{equation}
By taking a sufficiently large $p$, this constraint becomes less restrictive than the FFT-based one
in \cref{eq:ozaki-scheme-conv-fft-alpha}.

To further relax this constraint, we can combine NTTs with two different moduli using the CRT.
To evaluate $\bm{x}^{(s)} \circledast \bm{y}^{(t)}$ using NTTs modulo $p_0$ and $p_1$, we first compute the residues of
$c_x^{(s)}\bm{x}^{(s)}$ and $c_y^{(t)}\bm{y}^{(t)}$ modulo $p_0$ and $p_1$:
\begin{align}
	\bm{x}^{(s)}_{0} & = c_x^{(s)}\bm{x}^{(s)} \bmod p_0, \quad
  \bm{x}^{(s)}_{1} = c_x^{(s)}\bm{x}^{(s)} \bmod p_1, \\
	\bm{y}^{(t)}_{0} & = c_y^{(t)}\bm{y}^{(t)} \bmod p_0, \quad
  \bm{y}^{(t)}_{1} = c_y^{(t)}\bm{y}^{(t)} \bmod p_1.
\end{align}
We then compute the cyclic convolutions modulo $p_0$ and $p_1$ using the NTT:
\begin{align}
	\bm{z}^{\prime (s, t)}_{0}
  & = \NTT^{-1}_{p_0}\!\left(
      \NTT_{p_0}\!\left(\bm{x}^{(s)}_{0}\right)\odot
      \NTT_{p_0}\!\left(\bm{y}^{(t)}_{0}\right)
    \right)\bmod p_0, \\
	\bm{z}^{\prime (s, t)}_{1}
  & = \NTT^{-1}_{p_1}\!\left(
      \NTT_{p_1}\!\left(\bm{x}^{(s)}_{1}\right)\odot
      \NTT_{p_1}\!\left(\bm{y}^{(t)}_{1}\right)
    \right)\bmod p_1.
\end{align}
Finally, we reconstruct $\bm{x}^{(s)} \circledast \bm{y}^{(t)}$ via the CRT as
\begin{equation}
	\bm{x}^{(s)} \circledast \bm{y}^{(t)}
  = \left.
  \left(
    \bm{z}^{\prime (s, t)}_{0}\,p_1\,p_1^{-1}
    + \bm{z}^{\prime (s, t)}_{1}\,p_0\,p_0^{-1}
  \right)
  \middle/
  \left(c_x^{(s)} c_y^{(t)}\right)
  \right.,
\end{equation}
where $p_1^{-1}$ denotes the multiplicative inverse of $p_1$ modulo $p_0$, and $p_0^{-1}$ denotes the multiplicative inverse of $p_0$ modulo $p_1$.
Since the numerator corresponds to an integer cyclic convolution modulo $p_0p_1$ and $\left\|c_x^{(s)}\bm{x}^{(s)} \circledast c_y^{(t)}\bm{y}^{(t)}\right\|_\infty \le n\left\|c_x^{(s)}\bm{x}^{(s)}\right\|_{\infty} \cdot \left\|c_y^{(t)}\bm{y}^{(t)}\right\|_{\infty}$, a sufficient condition for exact recovery is
\begin{equation}
  n\left\|c_x^{(s)}\bm{x}^{(s)}\right\|_{\infty} \cdot
  \left\|c_y^{(t)}\bm{y}^{(t)}\right\|_{\infty}
  < \frac{p_0 p_1}{2}.
\end{equation}
Therefore, as in \cref{eq:ozaki-scheme-conv-ntt-alpha}, a suitable choice of $\alpha$ is
\begin{equation}
	\alpha
  = \left\lfloor \frac{\log_2\left(p_0 p_1 / 2\right) - \log_2\left(n\right)}{2} \right\rfloor.
  \label{eq:ozaki-scheme-conv-ntt-crt-alpha}
\end{equation}

\begin{figure}[t]
	\centering
	\includegraphics[width=\linewidth]{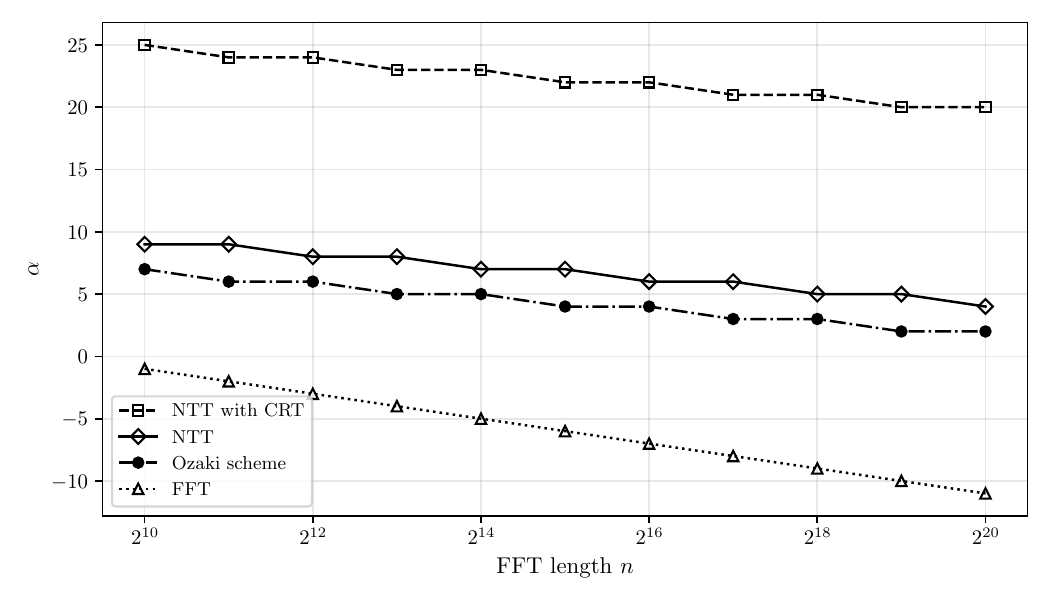}
	\caption{Comparison of split width $\alpha$ (bits per split component) for various convolution methods in single precision. A larger $\alpha$ implies fewer splits and thus lower computational cost.}
	\Description{Line plots of $\alpha$ (bits per split component) versus FFT length $n$ ($2^{10}$ to $2^{20}$) for four methods: original Ozaki scheme, FFT-based convolution, single-NTT, and CRT-based approach. The CRT-based $\alpha$ reaches 20 bits at $n=2^{20}$, substantially outperforming the other methods. The FFT-based $\alpha$ becomes negative for $n \leq 2^{10}$.}
	\label{fig:alpha-comparison}
\end{figure}

\Cref{fig:alpha-comparison} compares the resulting values of $\alpha$ for the considered methods in single precision.
In this setting, $u=2^{-24}$.
We also set $p=2{,}130{,}706{,}433 < 2^{31}$ in \cref{eq:ozaki-scheme-conv-ntt-alpha}, and $p_0=p$ and $p_1=2{,}113{,}929{,}217 < 2^{31}$ in \cref{eq:ozaki-scheme-conv-ntt-crt-alpha}.
When using FFT-based convolution, $\alpha$ in \cref{eq:ozaki-scheme-conv-fft-alpha} becomes negative for $n \ge 2^{10}$, meaning that not even one bit can be allocated.
Moreover, for the original Ozaki scheme, $\alpha$ in \cref{eq:ozaki-scheme-alpha} is only $2$ bits at $n=2^{20}$, suggesting that the number of splits may become large.
Using a single-modulus NTT, $\alpha$ in \cref{eq:ozaki-scheme-conv-ntt-alpha} is slightly larger than that of the original Ozaki scheme, but the improvement is modest.
In contrast, when using NTTs with two distinct moduli and reconstructing via the CRT, $\alpha$ in \cref{eq:ozaki-scheme-conv-ntt-crt-alpha} reaches $20$ bits at $n=2^{20}$, which is substantially larger than that of the original Ozaki scheme.
By \cref{eq:k-upper-limit}, a larger $\alpha$ implies fewer splits, and thus we can expect a reduction in computational cost.

\subsection{TS-Precision Cyclic Convolution with Ozaki Scheme}
Based on the above discussion, \cref{alg:ts-ozaki-scheme-convolution} presents a cyclic convolution algorithm for TS vectors using the Ozaki scheme.
Given TS vectors $\bm{x}_{0:2}$ and $\bm{y}_{0:2}$, it computes the cyclic convolution
$\bm{z}_{0:2} \simeq \bm{x}_{0:2}\circledast \bm{y}_{0:2}$
in the following three steps:
\begin{enumerate}[label=Step~\arabic*., align=left, labelindent=\parindent, leftmargin=!]
	\item Apply \cref{alg:ts-split} to $\bm{x}_{0:2}$ and $\bm{y}_{0:2}$ to split them into components and compute the 32-bit NTTs of each component modulo $p_0$ and $p_1$.
	\item For each pair of split components, compute their cyclic convolution via 32-bit inverse NTTs and CRT reconstruction.
	\item Sum all pairwise convolution results to obtain $\bm{z}_{0:2}$.
\end{enumerate}
\begin{algorithm}[t]
	\caption{TS-precision cyclic convolution with Ozaki scheme}
	\label{alg:ts-ozaki-scheme-convolution}
	\begin{algorithmic}[1]
		\Require TS vectors $\bm{x}_{0:2}, \bm{y}_{0:2}$
		\Require Distinct prime moduli $p_0, p_1 < 2^{32}$
		\Ensure TS-vector cyclic convolution $\bm{z}_{0:2} \simeq \bm{x}_{0:2} \circledast \bm{y}_{0:2}$
		\State $\bm{X}^{(0)}_0, \ldots, \bm{X}^{(k_x - 1)}_0, \bm{X}^{(0)}_1, \ldots, \bm{X}^{(k_x - 1)}_1, \bm{c}_x \gets \mathrm{SplitTSAndCalcNTT}\left(\bm{x}_{0:2}, p_0, p_1\right)$
		\State $\bm{Y}^{(0)}_0, \ldots, \bm{Y}^{(k_y - 1)}_0, \bm{Y}^{(0)}_1, \ldots, \bm{Y}^{(k_y - 1)}_1, \bm{c}_y \gets \mathrm{SplitTSAndCalcNTT}\left(\bm{y}_{0:2}, p_0, p_1\right)$
		\For {$s = 0, \ldots, k_x - 1$}
		\For {$t = 0, \ldots, k_y - 1$}
		\State $\bm{z}_{0}^{\prime (s, t)} \gets \NTT_{p_0}^{-1}\left(\bm{X}^{(s)}_{0} \odot \bm{Y}^{(t)}_{0}\right)$; $\bm{z}_{1}^{\prime (s, t)} \gets \NTT_{p_1}^{-1}\left(\bm{X}^{(s)}_{1} \odot \bm{Y}^{(t)}_{1}\right)$
		\State $\bm{z}'^{(s, t)} \gets \bm{z}_{0}^{\prime(s, t)} + \left\{\left(\bm{z}_{1}^{\prime(s, t)} - \bm{z}_{0}^{\prime(s, t)}\right) p_0^{-1} \bmod p_1\right\} p_0$
		\State $\bm{z}^{(s, t)}_{0:2} \gets \left. \bm{z}'^{(s, t)} \middle/ \left(c_x^{(s)} c_y^{(t)}\right) \right.$
		\EndFor
		\EndFor
		\State $\bm{z}_{0:2} \gets \bm{0}$
		\For {$s = 0, \ldots, k_x - 1$}
		\For {$t = 0, \ldots, k_y - 1$}
		\State $\bm{z}_{0:2} \gets \mathrm{TSAdd}\left(\bm{z}_{0:2}, \bm{z}^{(s, t)}_{0:2}\right)$
		\EndFor
		\EndFor
	\end{algorithmic}
\end{algorithm}
\begin{algorithm}[t]
	\caption{Splitting procedure and NTT computation in \cref{alg:ts-ozaki-scheme-convolution}}
	\label{alg:ts-split}
	\begin{algorithmic}[1]
		\Require TS vector $\bm{x}_{0:2}$
		\Require Distinct prime moduli $p_0, p_1 < 2^{32}$
		\Ensure $\bm{X}^{(0)}_0, \ldots, \bm{X}^{(k - 1)}_0 \in \mathbb{Z}_{p_0}^{n}$, $\bm{X}^{(0)}_1, \ldots, \bm{X}^{(k - 1)}_1 \in \mathbb{Z}_{p_1}^{n}$,  $\bm{c} \in \mathbb{F}^{k}_{32}$
		\Function {$\mathrm{SplitTSAndCalcNTT}~$}{$\bm{x}_{0:2}, p_0, p_1$}
		\State $\rho \gets \log_2 \left(u^{-1}_{32}\right) - \left\lfloor\left( \log_2 (p_0 p_1/2) - \log_2 (n)\right) / 2 \right\rfloor$
		\State $\bm{r}^{(0)}_{0:2} \gets \bm{x}_{0:2}$
		\State $\mu^{(0)} \gets \left\|\bm{r}^{(0)}_0\right\|_\infty$
		\State $k \gets 0$
		\While {$\mu^{(k)} > 0$}
		\State $\sigma^{(k)} \gets 2^{\left\lceil \log_2\left(\mu^{(k)}\right) \right\rceil + \rho}$
		\State $\bm{x}^{(k)} \gets \mathbf{fl}_{32} \left(\left(\bm{r}^{(k)}_{0} + \sigma^{(k)} \bm{1}_n \right) - \sigma^{(k)} \bm{1}_n\right)$
		\State $\bm{r}^{(k+1)}_{0}, \bm{r}^{(k+1)}_{1} \gets \mathrm{FastTwoSum}\left(\bm{r}^{(k)}_{0} -\bm{x}^{(k)}, \bm{r}^{(k)}_{1}\right)$
		\State $\bm{r}^{(k+1)}_{1}, \bm{r}^{(k+1)}_{2} \gets \mathrm{FastTwoSum}\left(\bm{r}^{(k)}_{1}, \bm{r}^{(k)}_{2}\right)$
		\State $c^{(k)} \gets \left(u_{32}\sigma^{(k)}\right)^{-1}$; $\bm{x}^{\prime (k)} \gets c^{(k)} \cdot \bm{x}^{(k)}$
		\State $\mu^{(k+1)} \gets \left\|\bm{r}^{(k+1)}_0\right\|_\infty$
		\State $\bm{X}^{(k)}_0 \gets \NTT_{p_0}\left(\bm{x}^{\prime (k)} \bmod p_0\right)$; $\bm{X}^{(k)}_1 \gets \NTT_{p_1}\left(\bm{x}^{\prime (k)} \bmod p_1\right)$
		\State $k \gets k + 1$
		\EndWhile
		\State \Return $\bm{X}^{(0)}_0, \ldots, \bm{X}^{(k - 1)}_0, \bm{X}^{(0)}_1, \ldots, \bm{X}^{(k - 1)}_1, \bm{c}$
		\EndFunction
	\end{algorithmic}
\end{algorithm}

In Step~1, applying \cref{alg:ts-split} to $\bm{x}_{0:2}$ and $\bm{y}_{0:2}$ yields
$\bm{X}^{(0)}_{0}, \ldots, \bm{X}^{(k_x - 1)}_{0}$,
$\bm{X}^{(0)}_{1}, \ldots, \bm{X}^{(k_x - 1)}_{1}$,
$\bm{c}_x$,
and similarly
$\bm{Y}^{(0)}_{0}, \ldots, \bm{Y}^{(k_y - 1)}_{0}$,
$\bm{Y}^{(0)}_{1}, \ldots, \bm{Y}^{(k_y - 1)}_{1}$,
$\bm{c}_y$.
Here, $p_0$ and $p_1$ are distinct primes with $p_0, p_1 < 2^{32}$ to enable 32-bit integer arithmetic,
and $p_0^{-1}$, $p_1^{-1}$ denote their multiplicative inverses satisfying
$p_0 p_0^{-1} \bmod p_1 = 1$ and $p_1 p_1^{-1} \bmod p_0 = 1$.
These quantities are constructed such that
\begin{align}
	\bm{x}_{0:2} & =
  \sum_{s = 0}^{k_x - 1}
  \left.
  \left(
    \NTT^{-1}_{p_0} \left(\bm{X}^{(s)}_{0}\right)p_1p_1^{-1}
    + \NTT^{-1}_{p_1} \left(\bm{X}^{(s)}_{1}\right)p_0p_0^{-1}
  \right)
  \middle/
  c_x^{(s)}
  \right.,
\end{align}
\begin{align}
	\bm{y}_{0:2} & =
  \sum_{t = 0}^{k_y - 1}
  \left.
  \left(
    \NTT^{-1}_{p_0} \left(\bm{Y}^{(t)}_{0}\right)p_1p_1^{-1}
    + \NTT^{-1}_{p_1} \left(\bm{Y}^{(t)}_{1}\right)p_0p_0^{-1}
  \right)
  \middle/
  c_y^{(t)}
  \right..
\end{align}

The splitting of a TS vector into single-precision vectors in \cref{alg:ts-split} follows the approach of the splitting algorithm proposed by Mukunoki et al.~\citeyearpar{Mukunoki2021} for binary128 vectors.
In \cref{alg:ts-split}, we first compute $\rho$ using $\alpha$ from \cref{eq:ozaki-scheme-conv-ntt-crt-alpha}:
\begin{align}
	\rho &= \log_2 \left(u_{32}^{-1}\right) - \alpha
  = \log_2 \left(u_{32}^{-1}\right) -
  \left \lfloor \frac{\log_2\left(p_0 p_1 / 2\right) - \log_2\left(n\right)}{2} \right \rfloor.
  \nonumber
\end{align}
We then iterate the following equations until $\mu^{(k)} = 0$
to obtain $\bm{X}^{(0)}_{0}, \ldots, \bm{X}^{(k - 1)}_{0}$,
$\bm{X}^{(0)}_{1}, \ldots, \bm{X}^{(k - 1)}_{1}$,
and $\bm{c}$:
\begin{align}
	\mu^{(k)} &= \left\|\bm{r}_{0}^{(k)}\right\|_{\infty} \label{eq:ts-split-mu} \\
	\sigma^{(k)} &= 2^{\left \lceil \log_2\left(\mu^{(k)}\right) \right \rceil + \rho} \nonumber \\
	\bm{r}^{(k)}_{0:2} &= \left\{
	\begin{array}{ll}
		\bm{x}_{0:2} & (k = 0) \\
		\bm{x}_{0:2} - \displaystyle \sum_{j = 0}^{k - 1} \bm{x}^{(j)} = \bm{r}^{(k - 1)}_{0:2} - \bm{x}^{(k - 1)} & (k > 0)
	\end{array} \label{eq:ts-split-r}
	\right. \\
	\bm{x}^{(k)} &= \mathbf{fl}_{32}\left(\left(\bm{r}^{(k)}_{0} + \sigma^{(k)} \bm{1}_n\right) - \sigma^{(k)} \bm{1}_n \right) \\
	c^{(k)} &= \left(u_{32} \sigma^{(k)}\right)^{-1} \\
	\bm{x}^{\prime(k)} &= c^{(k)} \cdot \bm{x}^{(k)} \\
	\bm{X}^{(k)}_{0} &= \NTT_{p_0}\!\left(\bm{x}^{\prime(k)} \bmod p_0\right) \nonumber\\ 
	\bm{X}^{(k)}_{1} &= \NTT_{p_1}\!\left(\bm{x}^{\prime(k)} \bmod p_1\right) \label{eq:ts-split-ntt-1}.
\end{align}
The subtraction of a single-precision number from a TS number in \cref{eq:ts-split-r} is performed using FastTwoSum~\cite{Dekker1971} (\cref{alg:fast_two_sum}).
FastTwoSum is an error-free transformation that computes the exact sum of two floating-point numbers $x$ and $y$ (with $|x| \ge |y|$) as a pair $(a, b)$ satisfying $s = \mathbf{fl}(x + y)$ and $a + b = x + y$.
\begin{algorithm}[t]
	\caption{FastTwoSum \cite{Dekker1971}}
	\label{alg:fast_two_sum}
	\begin{algorithmic}[1]
		\Require $x, y \in \mathbb{F}$ such that $\left|x\right| \ge \left|y\right|$
		\Ensure $a, b \in \mathbb{F}$ such that $a + b = x + y$
		\Function {$\mathrm{FastTwoSum}$}{$x, y$}
		\State $a \gets \mathbf{fl}(x + y)$
		\State $b \gets \mathbf{fl}(y - (a - x))$
		\State \Return $a, b$
		\EndFunction
	\end{algorithmic}
\end{algorithm}

In Step~2, for each $(s, t)$, we compute 32-bit inverse NTTs and then perform reconstruction via the CRT (in Garner form) to obtain the cyclic convolution of the split components:
\begin{align}
	\bm{z}^{\prime (s, t)}_{0} & = \NTT_{p_0}^{-1}\left(\bm{X}^{(s)}_{0} \odot \bm{Y}^{(t)}_{0}\right), \\
	\bm{z}^{\prime (s, t)}_{1} & = \NTT_{p_1}^{-1}\left(\bm{X}^{(s)}_{1} \odot \bm{Y}^{(t)}_{1}\right), \\
	\bm{z}^{\prime (s, t)} & =
  \bm{z}_{0}^{\prime(s, t)}
  + \left\{
    \left(\bm{z}_{1}^{\prime(s, t)} - \bm{z}_{0}^{\prime(s, t)}\right) p_0^{-1} \bmod p_1
  \right\} p_0, \\
	\bm{z}^{(s, t)}_{0:2} & =
  \left. \bm{z}^{\prime (s, t)} \middle/ \left(c_x^{(s)} c_y^{(t)}\right) \right..
\end{align}

Finally, in Step~3, we sum all pairwise convolution results to obtain $\bm{z}_{0:2}$:
\begin{align}
	\bm{z}_{0:2} = \sum_{s = 0}^{k_x - 1} \sum_{t = 0}^{k_y - 1} \bm{z}^{(s, t)}_{0:2}.
\end{align}

We now analyze the time complexity of \cref{alg:ts-ozaki-scheme-convolution}.
In Step~1, we perform $2\left(k_x+k_y\right)$ forward 32-bit NTTs.
Since each NTT of length $n$ requires $O(n\log n)$ time, Step~1 requires $O((k_x+k_y)n\log n)$ time in total.
In Step~2, we perform $2k_xk_y$ 32-bit inverse NTTs.
Since each inverse NTT requires $O(n\log n)$ time, Step~2 requires
$O\left(k_xk_yn\log n\right)$ time.
In Step~3, we add $k_xk_y$ vectors of length $n$, which requires
$O\left(k_xk_yn\right)$ time.
Therefore, Step~2 is the dominant term and the overall time complexity of
\cref{alg:ts-ozaki-scheme-convolution} is
$O\left(k_xk_yn\log n\right)$.
Moreover, the total number of 32-bit NTTs (including inverse NTTs) is
$2(k_x+k_y) + 2k_xk_y = 2(k_x+k_y+k_xk_y)$.

\subsection{Reduction of Computational Cost}
In \cref{alg:ts-split}, we split a TS vector until $\mu^{(k)}$ becomes zero. In \cref{alg:ts-ozaki-scheme-convolution}, we then compute the cyclic convolution for every pair of split components and obtain $\bm{z}_{0:2}$ by summing all the resulting terms.
In practice, however, sufficiently small terms $\bm{z}^{(s,t)}_{0:2}$ can be omitted from the sum with little loss of accuracy.
Therefore, to reduce both the number of forward NTTs in \cref{alg:ts-split} and the number of inverse NTTs in \cref{alg:ts-ozaki-scheme-convolution}, we impose an upper bound on the number of splits and terminate the splitting procedure early.
A similar strategy is employed in the Ozaki scheme for matrix multiplication to reduce the number of matrix--matrix products.

To further reduce the number of inverse NTTs, consider indices $s,t,s',t'$ satisfying
$c_x^{(s)} c_y^{(t)} = c_x^{(s')} c_y^{(t')}$.
In this case, the following identity holds:
\begin{align}
	\bm{z}_{0:2}^{(s, t)} + \bm{z}_{0:2}^{(s', t')} =& \left. \bm{z}^{\prime (s, t)} \middle/ \left(c_x^{(s)} c_y^{(t)}\right) \right. + \left. \bm{z}^{\prime (s', t')} \middle/ \left(c_x^{(s')} c_y^{(t')}\right) \right. \nonumber \\
	=& \left. \left(\bm{z}^{\prime (s, t)} + \bm{z}^{\prime (s', t')} \right) \middle/ \left(c_x^{(s)} c_y^{(t)}\right) \right. \nonumber \\
	=& \left. \left\{\left(\bm{z}_{0}^{\prime(s, t)} + \bm{z}_{0}^{\prime(s', t')}\right) p_1 p_1^{-1} + \left(\bm{z}_{1}^{\prime(s, t)} + \bm{z}_{1}^{\prime(s', t')}\right) p_0 p_0^{-1}\right\} \middle/ \left(c_x^{(s)} c_y^{(t)}\right) \right. \nonumber \\
	=& \left\{\left(\NTT_{p_0}^{-1}\left(\bm{X}^{(s)}_{0} \odot \bm{Y}^{(t)}_{0}\right) + \NTT_{p_0}^{-1}\left(\bm{X}^{(s')}_{0} \odot \bm{Y}^{(t')}_{0}\right)\right) p_1 p_1^{-1} \right. \nonumber \\
	&+ \left. \left. \left(\NTT_{p_1}^{-1}\left(\bm{X}^{(s)}_{1} \odot \bm{Y}^{(t)}_{1}\right) + \NTT_{p_1}^{-1}\left(\bm{X}^{(s')}_{1} \odot \bm{Y}^{(t')}_{1}\right)\right) p_0 p_0^{-1}\right\} \middle/ \left(c_x^{(s)} c_y^{(t)}\right) \right. \nonumber \\
	=& \left\{\left(\NTT_{p_0}^{-1}\left(\bm{X}^{(s)}_{0} \odot \bm{Y}^{(t)}_{0} + \bm{X}^{(s')}_{0} \odot \bm{Y}^{(t')}_{0}\right)\right) p_1 p_1^{-1} \right. \nonumber \\
	&+ \left. \left. \left(\NTT_{p_1}^{-1}\left(\bm{X}^{(s)}_{1} \odot \bm{Y}^{(t)}_{1} + \bm{X}^{(s')}_{1} \odot \bm{Y}^{(t')}_{1}\right)\right) p_0 p_0^{-1}\right\} \middle/ \left(c_x^{(s)} c_y^{(t)}\right) \right. .
\end{align}
Hence, when $c_x^{(s)} c_y^{(t)} = c_x^{(s')} c_y^{(t')}$, we can reduce the number of inverse NTTs by accumulating
$\bm{X}^{(s)}_{0} \odot \bm{Y}^{(t)}_{0}$ and $\bm{X}^{(s')}_{0} \odot \bm{Y}^{(t')}_{0}$, and
$\bm{X}^{(s)}_{1} \odot \bm{Y}^{(t)}_{1}$ and $\bm{X}^{(s')}_{1} \odot \bm{Y}^{(t')}_{1}$,
in the NTT domain before applying the inverse NTTs, rather than computing $\bm{z}_{0:2}^{(s, t)}$ and $\bm{z}_{0:2}^{(s', t')}$ separately and summing them afterward.

The equality $c_x^{(s)} c_y^{(t)} = c_x^{(s')} c_y^{(t')}$ holds, for example, when $s+t=s'+t'$.
Specifically, if the splitting proceeds without gaps, then
$c_x^{(1)} = 2^{-\alpha}c_x^{(0)}$ and $c_x^{(2)} = 2^{-2\alpha}c_x^{(0)}$,
and similarly for $c_y^{(1)}$ and $c_y^{(2)}$.
In that case, setting $s+t=s'+t'=k$ gives
$c_x^{(s)} c_y^{(t)} = c_x^{(s')} c_y^{(t')} = 2^{-k\alpha} c_x^{(0)} c_y^{(0)}$.

On the other hand, accumulating terms before the inverse NTTs may cause the integer cyclic convolution result to exceed the range
$[-p_0p_1/2,\; p_0p_1/2)$, in which case CRT reconstruction will fail.
To prevent this, we impose an upper bound $L$ on the number of terms accumulated before applying the inverse NTTs and choose $\alpha$ such that reconstruction remains valid even after at most $L$ additions:
\begin{align}
	\alpha = \left\lfloor \frac{\log_2(p_0p_1/2) - \log_2(Ln)}{2} \right\rfloor.
\end{align}
We accordingly perform the splitting using this value of $\alpha$.

\cref{alg:ts-ozaki-scheme-convolution-2} presents a modified version of \cref{alg:ts-ozaki-scheme-convolution} that incorporates this NTT-domain accumulation strategy.
Here, $\mathrm{SplitTSAndCalcNTT}(\bm{x}_{0:2}, p_0, p_1, L)$ denotes the splitting procedure that uses
\begin{align}
	\rho
  = \log_2 \left(u_{32}^{-1}\right) - \alpha
  = \log_2 \left(u_{32}^{-1}\right) - \left\lfloor \frac{\log_2(p_0 p_1 / 2) - \log_2(Ln)}{2} \right\rfloor
\end{align}
in place of $\rho$ in \cref{alg:ts-split}.
\begin{algorithm}[t]
	\caption{TS-precision cyclic convolution with Ozaki scheme and NTT-domain accumulation}
	\label{alg:ts-ozaki-scheme-convolution-2}
	\begin{algorithmic}[1]
		\Require TS vectors $\bm{x}_{0:2}, \bm{y}_{0:2}$
		\Require Distinct prime moduli $p_0, p_1 < 2^{32}$
		\Require Accumulation limit $L \in \mathbb{N}$
		\Ensure TS vectors $\bm{z}_{0:2} \simeq \bm{x}_{0:2} \circledast \bm{y}_{0:2}$
		\State $\bm{X}^{(0)}_0, \ldots, \bm{X}^{(k_x - 1)}_0, \bm{X}^{(0)}_1, \ldots, \bm{X}^{(k_x - 1)}_1, \bm{c}_x \gets \mathrm{SplitTSAndCalcNTT}\left(\bm{x}_{0:2}, p_0, p_1, L\right)$
		\State $\bm{Y}^{(0)}_0, \ldots, \bm{Y}^{(k_y - 1)}_0, \bm{Y}^{(0)}_1, \ldots, \bm{Y}^{(k_y - 1)}_1, \bm{c}_y \gets \mathrm{SplitTSAndCalcNTT}\left(\bm{y}_{0:2}, p_0, p_1, L\right)$
		\State $\bm{z}_{0:2}, \bm{Z}_{0}^{\prime}, \bm{Z}_{1}^{\prime} \gets \bm{0}$
		\State $c \gets c_x^{(k_x - 1)} \cdot c_y^{(k_y - 1)}$
		\State $l \gets 0$
		\For {$k = k_x + k_y - 2, \ldots, 0$}
		\For {$s = \max\left(0, k - k_y + 1\right), \ldots, \min \left(k_x - 1, k\right)$}
		\State $t \gets k - s$
		\If {$l \geq L$ or $c \neq c_x^{(s)} \cdot c_y^{(t)}$}
		\State $\bm{z}^{\prime}_{0} \gets \mathrm{NTT}^{-1}_{p_0} \left(\bm{Z}_{0}^{\prime}\right)$; $\bm{z}^{\prime}_{1} \gets \mathrm{NTT}^{-1}_{p_1} \left(\bm{Z}_{1}^{\prime}\right)$
		\State $\bm{z}^{\prime} \gets \bm{z}^{\prime}_{0} + \left\{\left(\bm{z}^{\prime}_{1} - \bm{z}^{\prime}_{0}\right) p_0^{-1} \bmod p_1\right\} p_0$
		\State $\bm{z}_{0:2}^{(k)} \gets \bm{z}^{\prime} / c$
		\State $\bm{z}_{0:2} \gets \mathrm{TSAdd}\left(\bm{z}_{0:2}, \bm{z}_{0:2}^{(k)}\right)$
		\State $\bm{Z}_{0}^{\prime}, \bm{Z}_{1}^{\prime} \gets \bm{0}$
		\State $c \gets c_x^{(s)} \cdot c_y^{(t)}$
		\State $l \gets 0$
		\EndIf
		\State $\bm{Z}_{0}^{\prime} \gets \left(\bm{Z}_{0}^{\prime} + \bm{X}^{(s)}_{0} \odot \bm{Y}^{(t)}_{0}\right) \bmod p_0$
		\State $\bm{Z}_{1}^{\prime} \gets \left(\bm{Z}_{1}^{\prime} + \bm{X}^{(s)}_{1} \odot \bm{Y}^{(t)}_{1}\right) \bmod p_1$
		\State $l \gets l + 1$
		\EndFor
		\EndFor
		\State $\bm{z}'_{0} \gets \mathrm{NTT}^{-1}_{p_0} \left(\bm{Z}_{0}^{\prime}\right)$; $\bm{z}'_{1} \gets \mathrm{NTT}^{-1}_{p_1} \left(\bm{Z}_{1}^{\prime}\right)$
		\State $\bm{z}' \gets \bm{z}'_{0} + \left\{\left(\bm{z}'_{1} - \bm{z}'_{0}\right) p_0^{-1} \bmod p_1\right\} p_0$
		\State $\bm{z}_{0:2}^{(0)} \gets \bm{z}' / c$
		\State $\bm{z}_{0:2} \gets \mathrm{TSAdd}\left(\bm{z}_{0:2}, \bm{z}_{0:2}^{(0)}\right)$
	\end{algorithmic}
\end{algorithm}

In \cref{alg:ts-ozaki-scheme-convolution-2}, for a fixed $k$, the inner loop body (line 7) executes at most $\min(k_x,k_y)$ times.
Therefore, if we cap the number of splits at $K$ and set the accumulation limit to satisfy $L \geq K$, then for every $k$ the number of inverse NTTs can be reduced to two.
Moreover, when $k_x=k_y=K$, the outer loop (line 6) runs $2K-1$ times, so the total number of inverse NTTs is at least $2(2K-1)=4K-2$.
Together with the $4K$ forward NTTs required during splitting, the total number of NTTs (including inverse NTTs) in \cref{alg:ts-ozaki-scheme-convolution-2} is at least $8K-2$.
In this setting, the best-case time complexity becomes $O(Kn\log n)$.

Setting $L > 1$ reduces $\alpha$ by approximately $\log_2(L)/2$~bits compared with that for $L = 1$.
Each split component carries at most $\alpha$ bits of information.
When $K$ is finite, setting $L > 1$ reduces the bit width per split component compared with that for $L = 1$ for the same $K$, potentially reducing the overall achievable precision.
Setting $K = \infty$ avoids this degradation, but the number of splits may become large, increasing the number of NTTs.
However, the NTT-domain accumulation strategy in \cref{alg:ts-ozaki-scheme-convolution-2} mitigates this.

\subsection{Double-Precision Bluestein FFT Using 32-bit NTTs}
We implement a double-precision Bluestein FFT based on TS arithmetic by computing the TS-precision complex cyclic convolution in \cref{alg:ts-bluestein-fft} using \cref{alg:ts-ozaki-scheme-convolution-2}, which relies only on 32-bit NTTs.

In \cref{alg:ts-bluestein-fft}, we evaluate the following TS-precision complex cyclic convolution:
\begin{align}
  \bm{x}^{\prime}_{0:2} \circledast \bm{\omega}^{*}_{0:2}
  &=
  \Re\left(\bm{x}^{\prime}_{0:2}\right) \circledast \Re\left(\bm{\omega}^{*}_{0:2}\right)
  - \Im\left(\bm{x}_{0:2}^{\prime}\right) \circledast \Im\left(\bm{\omega}_{0:2}^{*}\right) \\
  &+ i \left(
    \Im\left(\bm{x}_{0:2}^{\prime}\right) \circledast \Re\left(\bm{\omega}_{0:2}^{*}\right)
    + \Re\left(\bm{x}_{0:2}^{\prime}\right) \circledast \Im\left(\bm{\omega}_{0:2}^{*}\right)
  \right).
\end{align}
Since the right-hand side consists of four real cyclic convolutions, this computation can be performed by calling \cref{alg:ts-ozaki-scheme-convolution-2} four times.
However, a straightforward implementation would introduce redundant splitting: each TS vector would be split twice, resulting in a total of eight calls to \cref{alg:ts-split}.
To avoid this redundancy, we cache the splitting results and reuse them across calls to \cref{alg:ts-ozaki-scheme-convolution-2}, reducing the number of splitting operations from eight to four.

With this caching strategy in place, we now estimate the total number of 32-bit transforms required by \cref{alg:ts-bluestein-fft}.
We split the four TS vectors
$\Re\left(\bm{x}^{\prime}_{0:2}\right)$, $\Im\left(\bm{x}^{\prime}_{0:2}\right)$, $\Re\left(\bm{\omega}^{*}_{0:2}\right)$, and $\Im\left(\bm{\omega}^{*}_{0:2}\right)$,
calling \cref{alg:ts-split} four times in total.
Let $K$ be the upper bound on the number of splits.
Each call to \cref{alg:ts-split} performs at most $2K$ forward 32-bit NTTs (one for each modulus), hence the total number of forward 32-bit NTTs is at most $8K$.

The number of inverse NTTs in \cref{alg:ts-ozaki-scheme-convolution-2} depends on the NTT-domain accumulation bound $L$ (the maximum number of terms accumulated before applying inverse NTTs).
In the worst case, when $L=1$ (no NTT-domain accumulation), a single real cyclic convolution requires at most $2K^2$ inverse 32-bit NTTs.
Because one complex cyclic convolution consists of four real cyclic convolutions, the total number of inverse 32-bit NTTs is at most $8K^2$.
Therefore, \cref{alg:ts-bluestein-fft} performs at most $8K + 8K^2$ 32-bit transforms in total (including both forward and inverse NTTs).

In the best case, when $L \ge K$, one call to \cref{alg:ts-ozaki-scheme-convolution-2} requires only $4K-2$ inverse 32-bit NTTs, giving a total of at least $8K + 4(4K - 2) = 24K - 8$ transforms.
Furthermore, since the twiddle factor vectors $\Re\left(\bm{\omega}^{*}_{0:2}\right)$ and $\Im\left(\bm{\omega}^{*}_{0:2}\right)$ are independent of the input, their splitting and the associated $4K$ forward NTTs can be precomputed offline, reducing the number of transforms required at runtime to at least $20K - 8$ in the best case and at most $4K + 8K^2$ in the worst case.

In summary, since each 32-bit NTT of length $n$ requires $O(n\log n)$ time and the overall time complexity of \cref{alg:ts-bluestein-fft} is $O(Kn\log n)$ in the best case and $O(K^2 n\log n)$ in the worst case.

\section{Experiments} \label{sec:experiments}

\subsection{Setup for Accuracy Evaluation}
The following settings are common to all accuracy experiments in this section.
For the proposed method, we used the prime moduli $p_0 = 2{,}130{,}706{,}433 < 2^{31}$ and $p_1 = 2{,}113{,}929{,}217 < 2^{31}$.
We computed FFTs of power-of-two lengths from $n = 2^{10}$ to $2^{18}$.

Input data were generated in double precision as
\begin{equation} \label{eq:input-generation}
	\left(\mathrm{rand} - 0.5 \right) \cdot \exp{\left(\phi \cdot \mathrm{randn}\right)},
\end{equation}
where $\mathrm{rand}$ is a random variable uniformly distributed on $(0,1]$ and $\mathrm{randn}$ is a standard normal random variable.
The parameter $\phi$ controls the range of exponents: when $\phi=0$, the input follows a uniform distribution on $(-0.5,0.5]$; as $\phi$ increases, more values cluster near zero while the spread of exponents increases.
We evaluated three settings: $\phi=0$, $1.0$, and $4.0$.

Using the FFT result computed using MPFR at 106-bit precision (twice the 53-bit significand of double precision) as a reference, we evaluated the maximum relative error and the relative $\ell_2$ error defined by
\begin{equation}
	\max_{0\leq j < n} \left\{ \max \left(\left|\frac{\Re\left(y'(j)\right) - \Re\left(y(j)\right)}{\Re\left(y(j)\right)}\right|, \left|\frac{\Im\left(y'(j)\right) - \Im\left(y(j)\right)}{\Im\left(y(j)\right)}\right|\right) \right\},
\end{equation}
and
\begin{equation}
	\frac{\|\bm{y}' - \bm{y}\|_2}{\|\bm{y}\|_2},
\end{equation}
where $\bm{y} \in \mathbb{C}^n$ is the MPFR (106-bit) reference output and $\bm{y}' \in \mathbb{C}^n$ is the output of the FFT method under evaluation.

\subsection{Accuracy Comparison with Other Double-Precision FFTs}
We compared the accuracy of the proposed double-precision Bluestein FFT (\cref{alg:ts-bluestein-fft} using \cref{alg:ts-ozaki-scheme-convolution-2}) with that of other double-precision FFT implementations.
As baselines, we considered the following five methods:
\begin{enumerate}[label=\arabic*.]
	\item Double-precision FFT using FFTW~\cite{Frigo2005}
	\item Stockham FFT using double-precision arithmetic
	\item Bluestein FFT using double-precision arithmetic
	\item Stockham FFT using TS arithmetic
	\item Bluestein FFT using TS arithmetic
\end{enumerate}
For methods~4 and~5, the input was converted from double precision to TS precision before computation, and the result was rounded back to double precision, as in the proposed method.
We set the upper bound on the number of splits to $K=\infty$ (no cap) and the NTT-domain accumulation bound to $L=1$ (no NTT-domain accumulation) for the proposed method.

\begin{figure}[t]
	\centering
	\includegraphics[width=0.9\hsize]{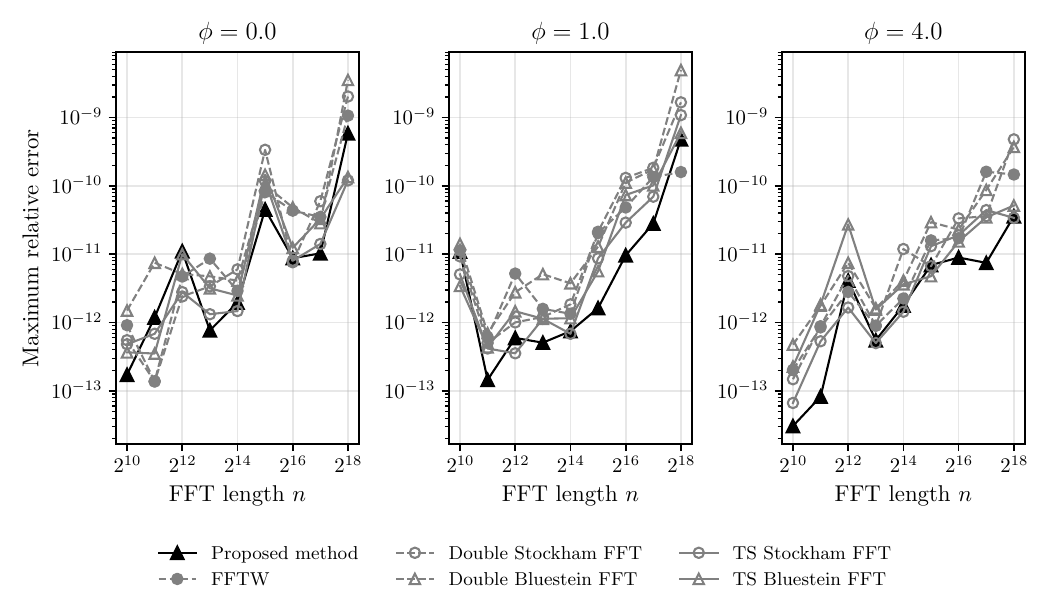}
	\caption{Maximum relative error of proposed double-precision Bluestein FFT ($(K,\,L)=(\infty,\,1)$) compared with that of other double-precision FFT implementations.}
	\Description{Line plots of maximum relative error versus FFT length ($2^{10}$ to $2^{18}$) for six methods and three values of $\phi$ (0, 1.0, 4.0). The proposed method does not consistently achieve the smallest maximum relative error.}
	\label{fig:max-rel-error-comparison-1}
\end{figure}

\Cref{fig:max-rel-error-comparison-1} shows the maximum relative error.
While the Ozaki scheme tends to reduce the maximum relative error in matrix multiplication~\cite{Ozaki2012}, the proposed method does not consistently achieve the smallest maximum relative error for FFTs.

\begin{figure}[t]
	\centering
	\includegraphics[width=0.9\hsize]{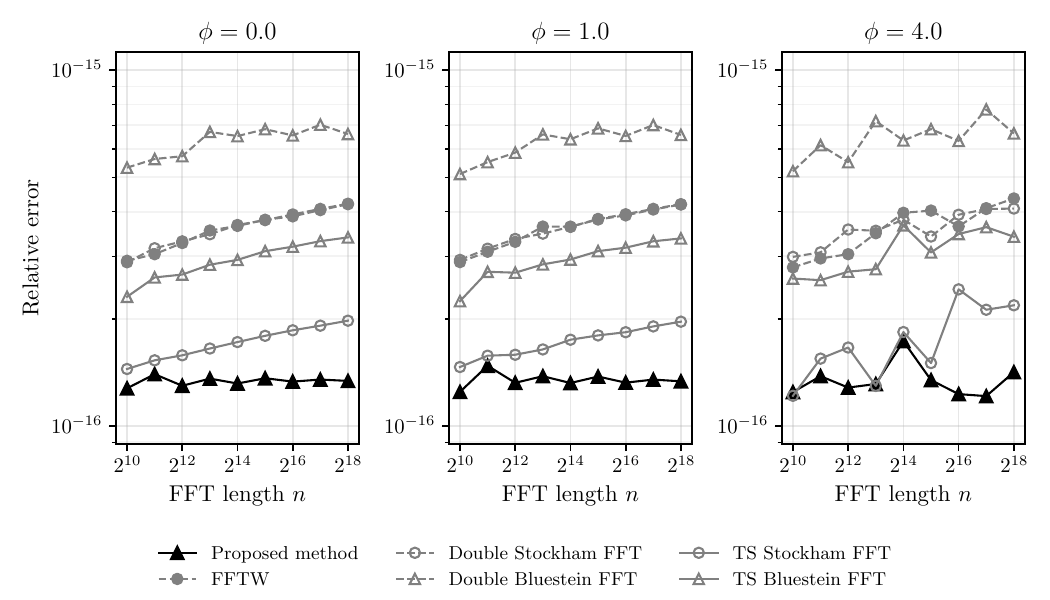}
	\caption{Relative error of proposed double-precision Bluestein FFT ($(K,\,L)=(\infty,\,1)$) compared with that of other double-precision FFT implementations.}
	\Description{Line plots of relative error versus FFT length for six methods and three values of $\phi$. The proposed method achieves the smallest relative error in most cases, with TS-based implementations outperforming their double-precision counterparts.}
	\label{fig:rel-error-comparison-1}
\end{figure}

\Cref{fig:rel-error-comparison-1} presents the relative error.
The Stockham FFTs exhibit smaller relative errors than those of the Bluestein FFTs, presumably because the Stockham FFT requires fewer arithmetic operations and hence accumulates less rounding error.
For a given FFT algorithm, TS-based implementations yield smaller relative errors than those of their double-precision counterparts, which can be attributed to the longer effective significand of the TS format.
Moreover, the proposed method achieves the smallest relative error in most cases; the only exceptions are for $(\phi, n) = (4.0,\, 2^{10})$ and $(4.0,\, 2^{13})$, where the TS-based Stockham FFT is marginally more accurate.
In addition, the proposed method substantially reduces the relative error compared with that of the TS-based Bluestein FFT, which differs only in the cyclic convolution routine.
These results indicate that applying the Ozaki scheme to the cyclic convolution stage is effective in reducing the relative error of FFTs.

For the baseline double-precision FFTs, the relative error tends to increase with $n$.
The proposed method does not exhibit such a trend.
In general, increasing $n$ increases the number of operations in an FFT and increases rounding error accumulation.
In contrast, in the proposed method, the Ozaki scheme is applied to the cyclic convolution in Step~3 of \cref{alg:ts-bluestein-fft}, which dominates the computational cost, and the split cyclic convolutions can be computed exactly; this likely suppresses the growth of relative error, unlike the other methods.

\begin{figure}[t]
	\centering
	\includegraphics[width=0.9\hsize]{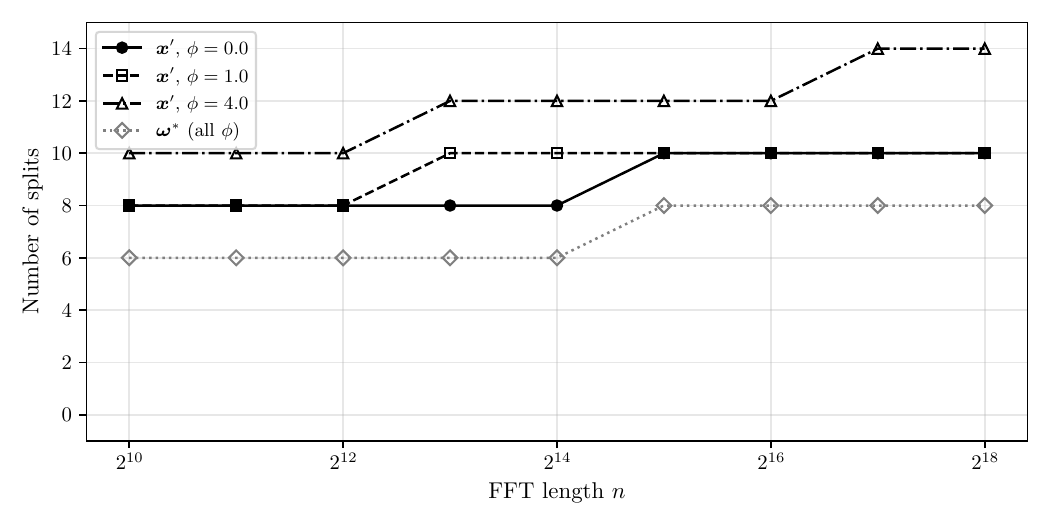}
	\caption{Number of splits of $\bm{x}'$ and $\bm{\omega}^{*}$ in proposed double-precision Bluestein FFT ($(K,\,L)=(\infty,\,1)$).}
	\Description{Line plots of the number of splits of $\bm{x}'$ and $\bm{\omega}^{*}$ versus FFT length for three values of $\phi$. The number of splits of $\bm{x}'$ increases with both $\phi$ and $n$, while that of $\bm{\omega}^{*}$ increases only with $n$.}
	\label{fig:split-size-comparison}
\end{figure}

\begin{figure}[t]
	\centering
	\includegraphics[width=0.9\hsize]{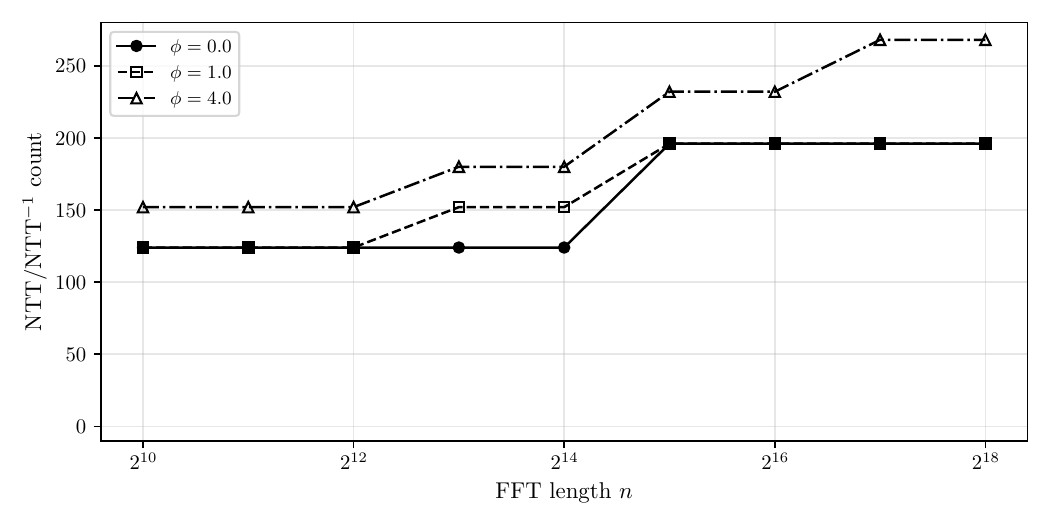}
	\caption{Total number of 32-bit NTTs and inverse NTTs in proposed double-precision Bluestein FFT ($(K,\,L)=(\infty,\,1)$).}
	\Description{Line plots of the total number of 32-bit NTTs and inverse NTTs versus FFT length for three values of $\phi$. The count increases with both $\phi$ and $n$, reaching a maximum of 268 at $\phi=4.0$ and $n=2^{18}$.}
	\label{fig:ntt-count-comparison}
\end{figure}

For this setting, the numbers of splits of $\bm{x}'$ and $\bm{\omega}^*$ and the numbers of 32-bit NTTs and inverse NTTs are shown in \cref{fig:split-size-comparison,fig:ntt-count-comparison}, respectively.
The number of splits of $\bm{x}'$ tends to increase as $\phi$ and $n$ increase.
In contrast, the number of splits of $\bm{\omega}^*$ increases with $n$ but is independent of $\phi$, since $\bm{\omega}^*$ does not depend on the input.
The total number of 32-bit NTTs and inverse NTTs also increases with $\phi$ and $n$; it is maximized at $(\phi, n) = (4.0, 2^{17})$ and $(4.0, 2^{18})$, where the total count is 268.

\subsection{Effect of Upper Bound on Number of Splits}
\begin{figure}[t]
	\centering
	\includegraphics[width=0.9\hsize]{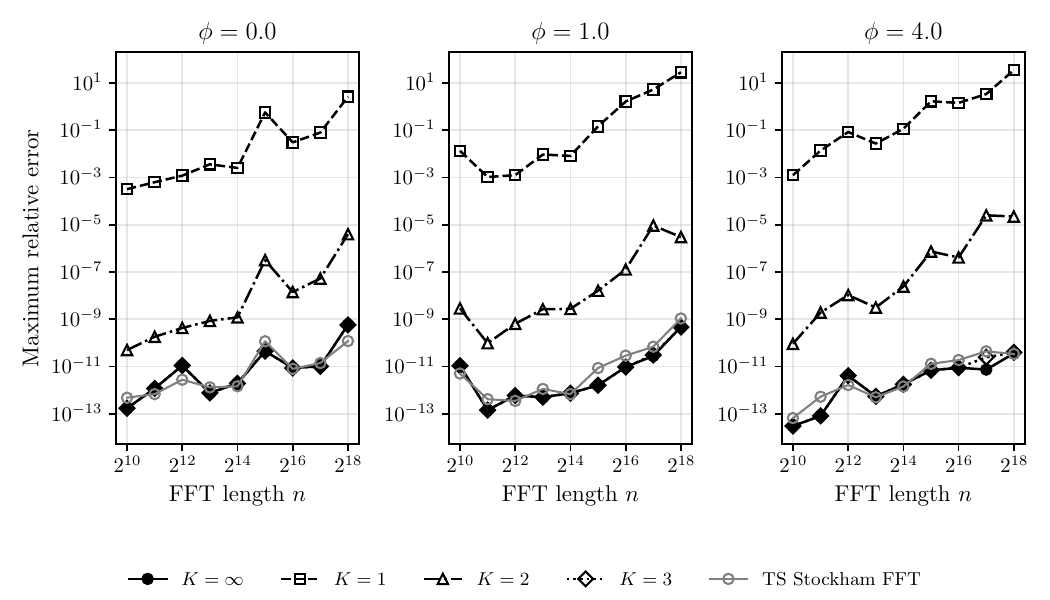}
	\caption{Maximum relative error of proposed double-precision Bluestein FFT ($L=1$) for various values of upper bound $K$ on number of splits.}
	\Description{Line plots of maximum relative error versus FFT length for $K=\infty$, 1, 2, 3 and the TS-based Stockham FFT for three values of $\phi$. Error decreases as $K$ increases.}
	\label{fig:max-rel-error-comparison-2}
\end{figure}

\begin{figure}[t]
	\centering
	\includegraphics[width=0.9\hsize]{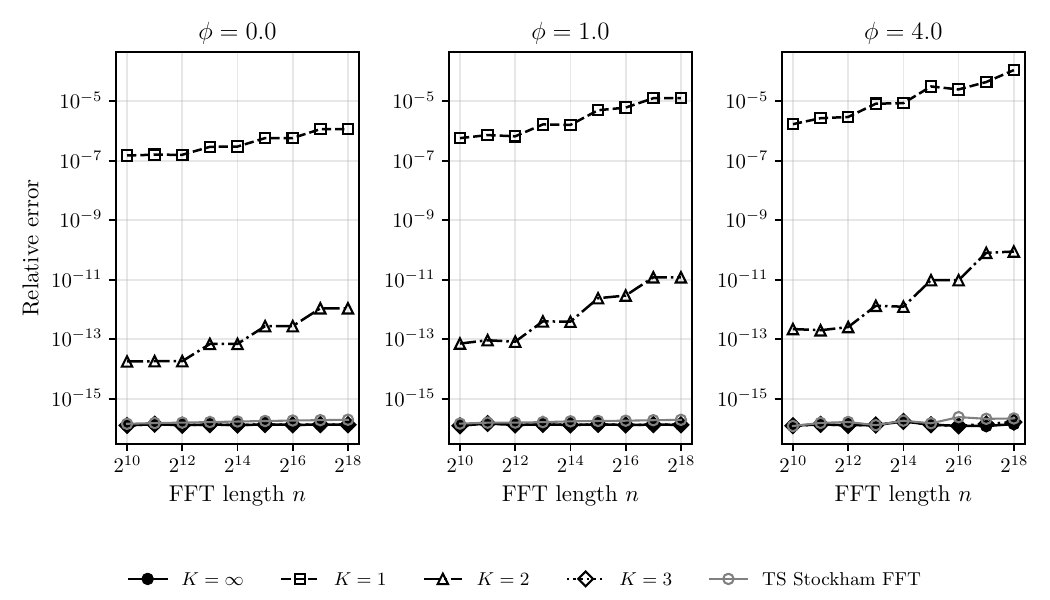}
	\caption{Relative error of proposed double-precision Bluestein FFT ($L=1$) for various values of upper bound $K$ on number of splits.}
	\Description{Line plots of relative error versus FFT length for $K=\infty$, 1, 2, 3 and the TS-based Stockham FFT for three values of $\phi$. $K=3$ achieves accuracy comparable to that for $K=\infty$.}
	\label{fig:rel-error-comparison-2}
\end{figure}

Next, we evaluated the maximum relative error and the relative error by varying the upper bound $K$ on the number of splits as $\infty, 1, 2, 3$.
The results are shown in \cref{fig:max-rel-error-comparison-2,fig:rel-error-comparison-2}, respectively.
For comparison, we also include results for the TS-based Stockham FFT.
Both error measures decrease as $K$ increases.
For $K=3$, we obtain accuracy comparable to that of the proposed method with $K=\infty$ and to that of the TS-based Stockham FFT.

\begin{figure}[t]
	\centering
	\includegraphics[width=0.9\hsize]{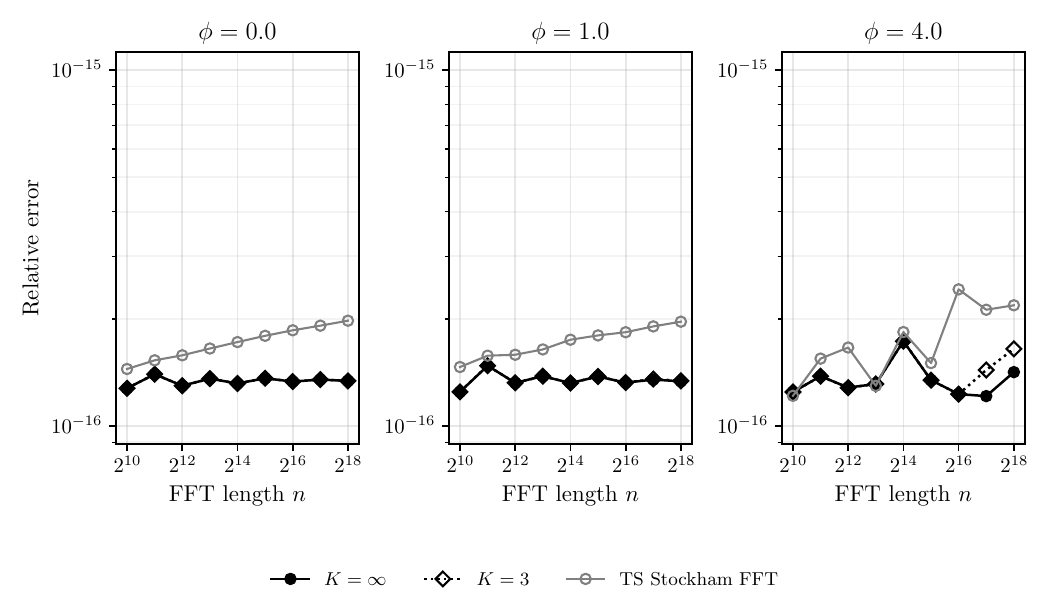}
	\caption{Enlarged view of \cref{fig:rel-error-comparison-2} ($L=1$) in range $[10^{-16}, 10^{-15}]$.}
	\Description{Enlarged line plots of relative error in the range $10^{-16}$ to $10^{-15}$, showing that $K=3$ achieves relative errors comparable to or smaller than those for the TS-based Stockham FFT for most combinations of $n$ and $\phi$.}
	\label{fig:rel-error-comparison-3}
\end{figure}

\Cref{fig:rel-error-comparison-3} shows an enlarged view of the range $[10^{-16}, 10^{-15}]$ in \cref{fig:rel-error-comparison-2}.
As can be seen, for most combinations of $n$ and $\phi$, $K=3$ achieves relative errors comparable to those achieved with $K=\infty$ and smaller than those of the TS-based Stockham FFT.
For $K=3$, the total number of 32-bit NTTs and inverse NTTs is 96, which is reduced to approximately 36\% of the maximum count of 268 for $K=\infty$.
This indicates that using a small upper bound $K$ on the number of splits is effective for reducing computational cost while maintaining accuracy.

\subsection{Effect of NTT-Domain Accumulation Before Inverse NTT}
\begin{figure}[t]
	\centering
	\includegraphics[width=0.9\hsize]{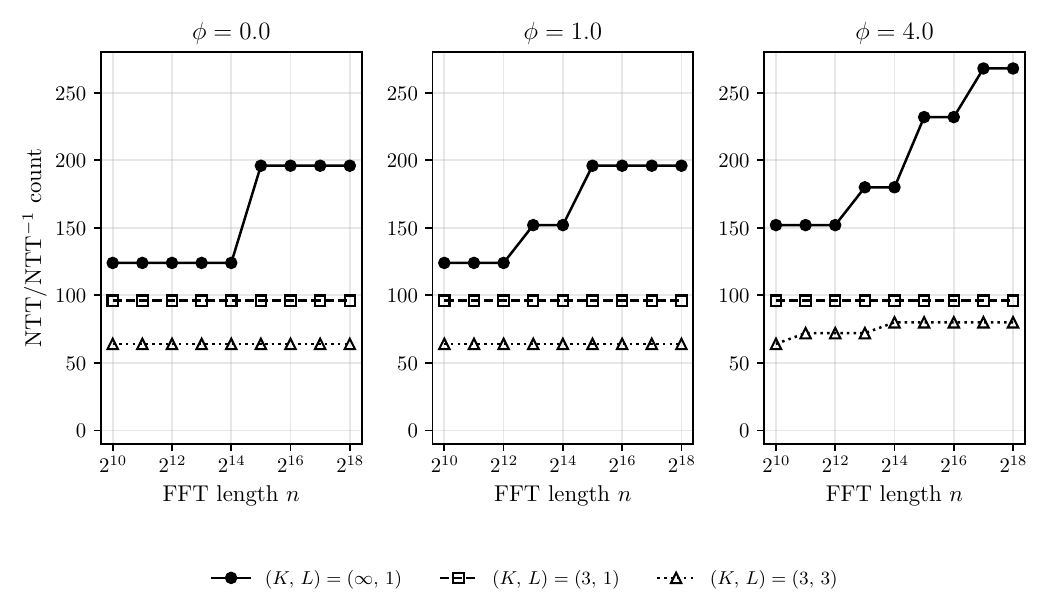}
	\caption{Effect of NTT-domain accumulation before inverse NTTs on total number of 32-bit NTTs and inverse NTTs in proposed double-precision Bluestein FFT.}
	\Description{Line plots comparing the total number of 32-bit NTTs and inverse NTTs for $K=3$, $L=3$ against $K=\infty$, $L=1$ and $K=3$, $L=1$ across FFT lengths and three values of $\phi$. NTT-domain accumulation reduces the NTT count, particularly for $\phi=0$ and 1.0.}
	\label{fig:ntt-count-comparison-2}
\end{figure}

We evaluated how NTT-domain accumulation before inverse NTTs reduces the number of 32-bit inverse NTTs.
\Cref{fig:ntt-count-comparison-2} compares the total numbers of 32-bit NTTs and inverse NTTs for $(K,\,L)=(3,\,3)$ against those for $(K,\,L)=(\infty,\,1)$ and $(3,\,1)$.
For $\phi=0$ and $1.0$, the setting $(K,\,L)=(3,\,3)$ yields 64 total 32-bit NTTs and inverse NTTs for all FFT sizes.
This matches the minimum count of 64 obtained by substituting $K=3$ into $24K - 8$.
Compared with the maximum count of 196 for $(K,\,L)=(\infty,\,1)$ at $\phi=0$ and $1.0$, this corresponds to a reduction to approximately 33\%.
For $\phi=4.0$, the maximum count for $(K,\,L)=(3,\,3)$ was 80; although this falls short of the minimum of 64, it is still a reduction to approximately 30\% of the maximum count of 268 for $(K,\,L)=(\infty,\,1)$.

\begin{figure}[t]
	\centering
	\includegraphics[width=0.9\hsize]{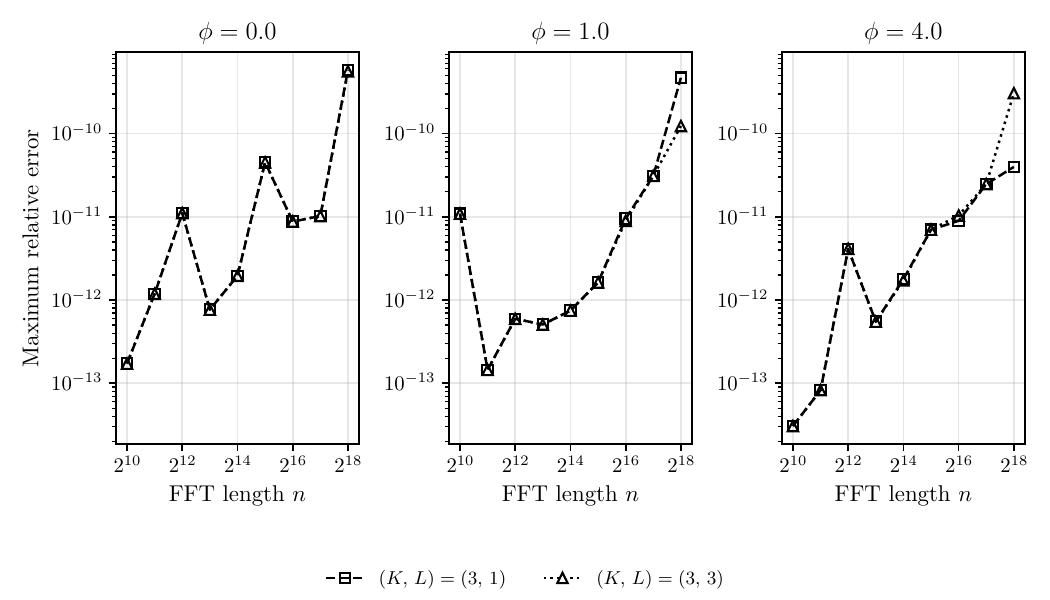}
	\caption{Maximum relative error of proposed double-precision Bluestein FFT ($K=3$) for various values of NTT-domain accumulation bound $L$ before inverse NTTs.}
	\Description{Line plots of maximum relative error versus FFT length for $K=3$ with $L=1$ and $L=3$ for three values of $\phi$. For $\phi=0$, $L=1$ and $L=3$ yield nearly identical errors. For $\phi=1.0$ and 4.0, $L=3$ yields larger errors than those for $L=1$.}
	\label{fig:max-rel-error-comparison-4}
\end{figure}

\begin{figure}[t]
	\centering
	\includegraphics[width=0.9\hsize]{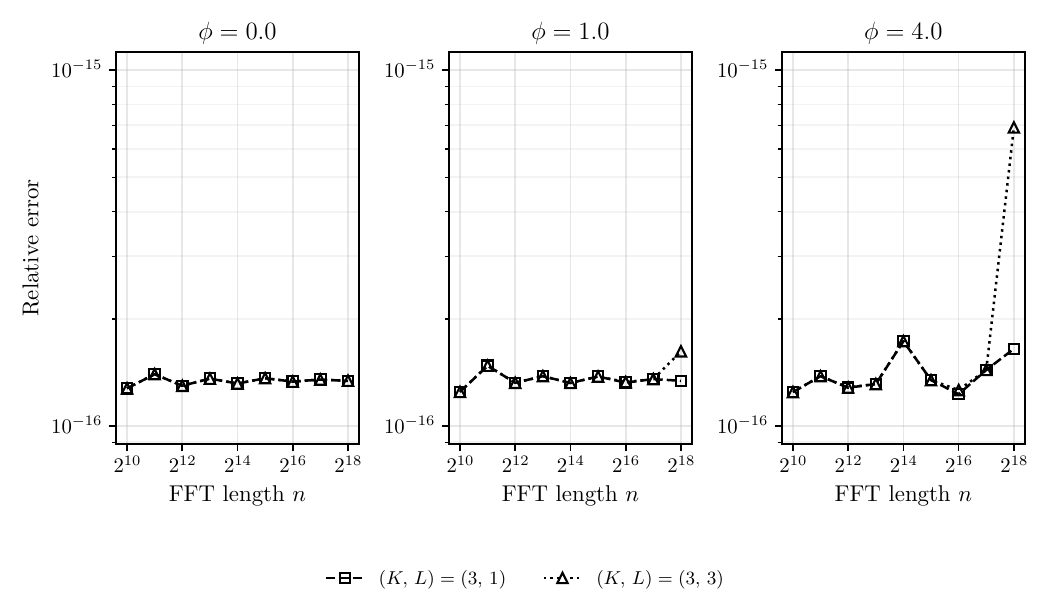}
	\caption{Relative error of proposed double-precision Bluestein FFT ($K=3$) for various values of NTT-domain accumulation bound $L$ before inverse NTTs.}
	\Description{Line plots of relative error versus FFT length for $K=3$ with $L=1$ and $L=3$ for three values of $\phi$. For $\phi=0$, $L=1$ and $L=3$ yield nearly identical errors. For $\phi=1.0$ and 4.0, $L=3$ yields larger errors than those for $L=1$.}
	\label{fig:rel-error-comparison-4}
\end{figure}

\Cref{fig:max-rel-error-comparison-4,fig:rel-error-comparison-4} compare the maximum relative error and relative error, respectively, for $(K,\,L)=(3,\,1)$ and $(3,\,3)$.
While no significant error difference is observed for $\phi = 0$, $(K,\,L) = (3,\,3)$ yields larger relative errors than does $(3,\,1)$ at $(\phi,\,n) = (1.0,\,2^{18})$ and $(4.0,\,2^{18})$.
For $n = 2^{17}$, both $(K,\,L) = (3,\,1)$ and $(3,\,3)$ yield $\alpha = 21$~bits, whereas for $n = 2^{18}$, $(K,\,L) = (3,\,1)$ still yields $\alpha = 21$~bits but $(3,\,3)$ yields $\alpha = 20$~bits, resulting in a total precision loss of $1 \times 3 = 3$~bits.

\Cref{fig:rel-error-comparison-4} reveals the precision required for accurate computation.
For $\phi = 1.0$ and $4.0$, in the range where $\alpha \geq 21$~bits with $L = 3$ ($n \leq 2^{17}$), the relative errors for $L = 3$ and $L = 1$ are nearly identical; however, once $\alpha$ drops to 20~bits ($n = 2^{18}$), a noticeable difference emerges.
This observation suggests that the total required bit width is at least $21 \times 3 = 63$~bits for $\phi = 1.0$ and $4.0$.

These results suggest that for inputs with a narrow exponent range, such as $\phi = 0$, NTT-domain accumulation is effective: the maximum number of terms is consistently accumulated before each inverse NTT call, with little accuracy loss.
However, the decrease in $\alpha$ caused by NTT-domain accumulation can adversely affect accuracy for inputs with a wide exponent range.
Moreover, since the maximum accumulation count is rarely reached in this case, the reduction in the number of inverse NTTs is also limited.

\subsection{Execution Time}
We measured the execution time of the proposed double-precision Bluestein FFT with $(K,\,L)=(3,\,3)$ and input data generated by \cref{eq:input-generation} with $\phi=0$.
FFT lengths ranged from $n = 2^{10}$ to $2^{18}$. 
The evaluation platform is described in \cref{tab:experimental-environment}.
The entire implementation, including the 32-bit NTT and TS arithmetic, was written from scratch in C++.
The 32-bit NTT is based on the Stockham algorithm.
Modular multiplications are performed using Montgomery multiplication~\cite{Montgomery1985} and Shoup multiplication~\cite{Faernqvist2005}.
Independent loop iterations were parallelized using OpenMP with 48 threads.
We did not use hand-written SIMD intrinsics or explicit cache optimization; vectorization relies solely on compiler auto-vectorization.
We performed 10 warm-up runs before measuring the execution time of a single run.
\begin{table}[t]
	\centering
	\caption{Experimental environment for execution time comparison.}
	\Description{Table listing the hardware and software specifications of the experimental environment.}
	\label{tab:experimental-environment}
	\begin{tabular}{ll}
		\toprule
		CPU            & Intel Xeon Platinum 8468 \\
		Cores/Threads  & 48 cores / 48 threads \\
		Memory         & 128 GB DDR5-4800 \\
		Compiler       & Intel(R) oneAPI C++ Compiler 2025.0.4 \\
		Compiler flags  & -O3 -xHOST -qopenmp -fma \\
		FFTW version   & 3.3.10 \\
		\bottomrule
	\end{tabular}
\end{table}

\begin{figure}[t]
	\centering
	\includegraphics[width=0.9\hsize]{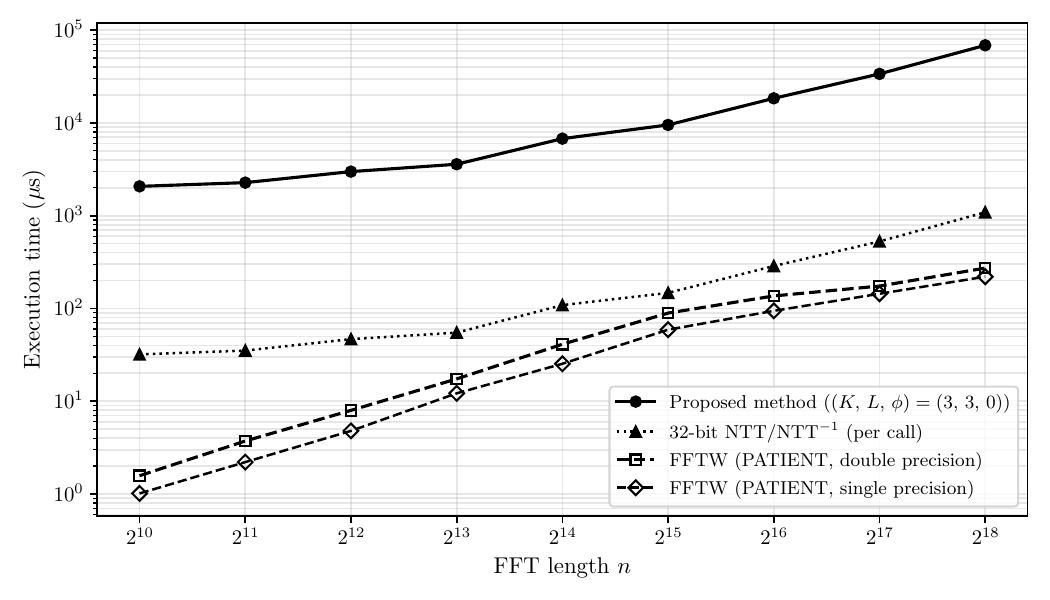}
	\caption{Execution time of proposed double-precision Bluestein FFT ($(K,\,L,\,\phi)=(3,\,3,\,0)$) compared with FFTW's double- and single-precision FFTs (PATIENT mode, 48 threads), together with per-call execution time of single 32-bit NTT/NTT$^{-1}$.}
	\Description{Log-log plot of execution time versus FFT length from $2^{10}$ to $2^{18}$ for the proposed method, FFTW (double precision), FFTW (single precision), and per-call 32-bit NTT/NTT$^{-1}$. The execution time of the proposed method is approximately 107--1315$\times$ that of FFTW double precision, and the per-call NTT time is comparable to FFTW single precision for large FFT lengths.}
	\label{fig:execution-time}
\end{figure}

\begin{figure}[t]
	\centering
	\includegraphics[width=0.9\hsize]{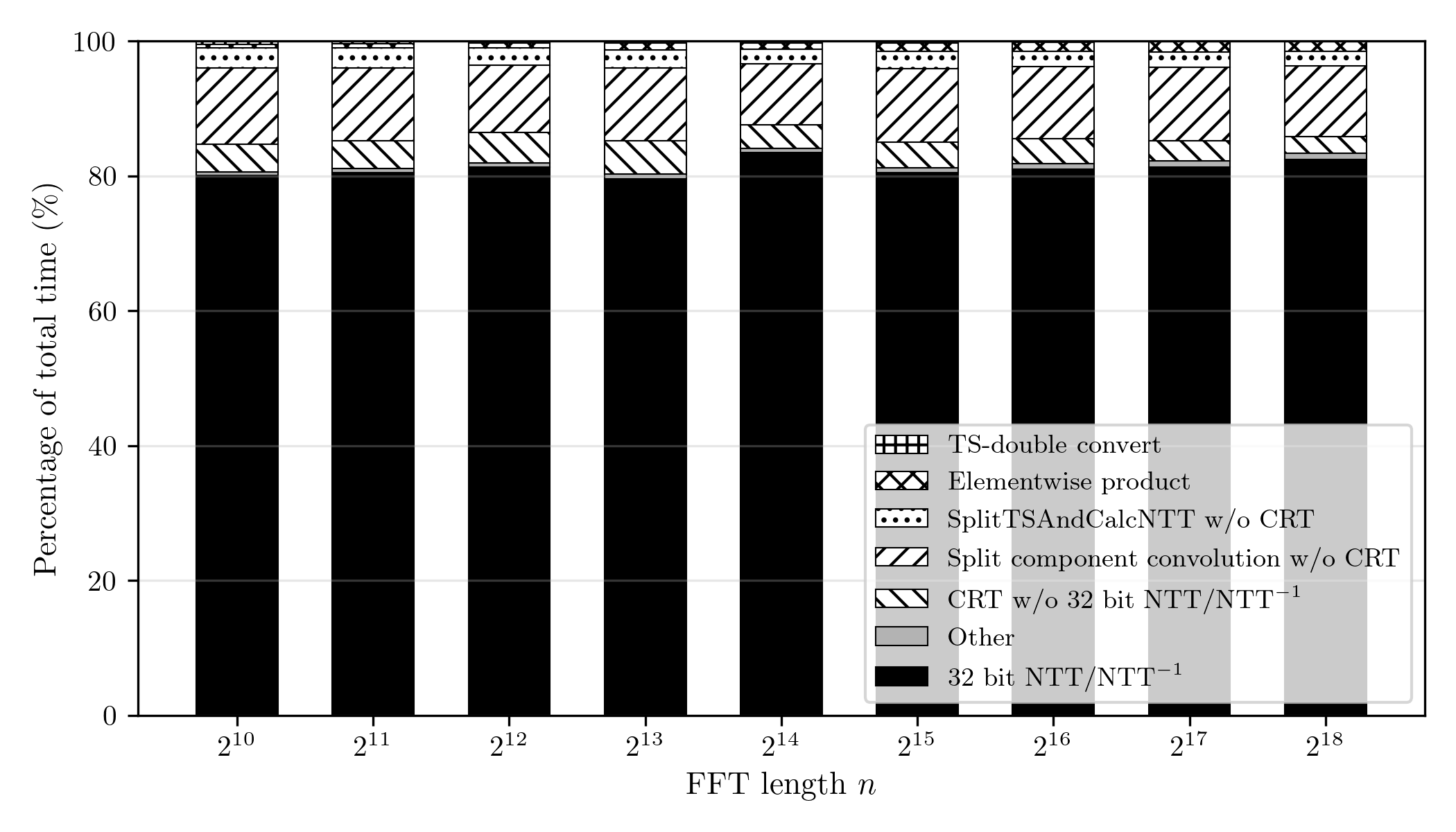}
	\caption{Execution time breakdown of proposed double-precision Bluestein FFT ($(K,\,L,\,\phi)=(3,\,3,\,0)$).}
	\Description{100\% stacked bar chart showing the execution time breakdown for FFT lengths from $2^{10}$ to $2^{18}$. The 32-bit NTT/NTT$^{-1}$ dominates at approximately 80\% of total time. The remaining time is distributed among split component convolution, CRT, elementwise product, split, TS-double conversion, and other operations.}
	\label{fig:time-breakdown}
\end{figure}

\Cref{fig:execution-time} shows the execution time.
For all values of $n$, the number of 32-bit NTT/NTT$^{-1}$ invocations was 52 (excluding the forward NTTs required for splitting the twiddle factor vector $\bm{\omega}^{*}$, which can be precomputed offline).
The figure also includes the per-call execution time of a single 32-bit NTT/NTT$^{-1}$, obtained by dividing the total NTT time by 52, as well as the execution time of double- and single-precision FFTs computed by FFTW (PATIENT mode, 48 threads, compiled with AVX-512 support) for reference.
The FFTW execution times were obtained by averaging 52 calls after 10 warm-up runs.
\Cref{fig:time-breakdown} shows the execution time breakdown of the proposed method for each FFT length.

The execution time of the proposed method is approximately 107 times ($n = 2^{15}$) to 1315 times ($n = 2^{10}$) that of FFTW's double-precision FFT, with the ratio being smallest around $n = 2^{15}$ and increasing for both smaller and larger $n$.

\Cref{fig:time-breakdown} shows that the 32-bit NTT/NTT$^{-1}$ accounts for approximately 80\% of the total execution time across all FFT lengths, while the remaining operations, namely CRT reconstruction, split component convolution, elementwise products, splitting, and data type conversion, each contribute a relatively small fraction.

Furthermore, the per-call execution time of the 32-bit NTT/NTT$^{-1}$ ranges from approximately $31$ times ($n = 2^{10}$) to $2.5$ times ($n = 2^{15}$) that of FFTW's single-precision FFT.
Although it is not guaranteed that the 32-bit NTT can be optimized to the level of FFTW's single-precision FFT, if we hypothetically assume that this is the case, the share of NTT in the total execution time would decrease from approximately 80\% to a range of 11\% ($n = 2^{10}$) to 62\% ($n = 2^{15}$).
In particular, for small $n$, the NTT would no longer dominate the total execution time, and even for large $n$, the non-NTT components cannot be neglected.
This suggests that reducing the overall execution time requires optimizing the non-NTT components as well.
Since the time complexity of the NTT computation is $O(Kn \log n)$ while that of the non-NTT components is $O(Kn)$, the ratio between them is only $O(\log n)$, which confirms that the non-NTT cost is not negligible.

\section{Discussion} \label{sec:discussion}
We now analyze the conditions under which the proposed method could achieve execution time that is competitive with a baseline double-precision FFT.
Let $T_{\mathrm{prop}}$ denote the total execution time of the proposed method,
$T_{\mathrm{NTT}}$ denote the per-call execution time of a single 32-bit NTT or NTT$^{-1}$,
$C_{\mathrm{NTT}}$ denote the total count of NTT/NTT$^{-1}$ invocations at runtime,
and $T_{\mathrm{base}}$ denote the execution time of the baseline double-precision FFT.
Let $P_{\mathrm{NTT}} = C_{\mathrm{NTT}} \cdot T_{\mathrm{NTT}} / T_{\mathrm{prop}}$ denote the fraction of $T_{\mathrm{prop}}$ spent on NTT computations.
From this definition, $T_{\mathrm{prop}} = C_{\mathrm{NTT}} \cdot T_{\mathrm{NTT}} / P_{\mathrm{NTT}}$ and the condition $T_{\mathrm{prop}} < T_{\mathrm{base}}$ is equivalent to
\begin{equation}
  \frac{T_{\mathrm{base}}}{T_{\mathrm{NTT}}} > \frac{C_{\mathrm{NTT}}}{P_{\mathrm{NTT}}}.
  \label{eq:speedup-condition}
\end{equation}

In our experiments with $(K,\,L)=(3,\,3)$, we have $C_{\mathrm{NTT}} = 52$ and $P_{\mathrm{NTT}} \approx 0.8$,
so \cref{eq:speedup-condition} requires $T_{\mathrm{base}} / T_{\mathrm{NTT}} > 65$.
If the NTT were optimized to run as fast as FFTW's single-precision FFT,
$P_{\mathrm{NTT}}$ would drop to approximately $0.62$ ($n = 2^{15}$) to $0.11$ ($n = 2^{10}$),
raising the threshold to $C_{\mathrm{NTT}} / P_{\mathrm{NTT}} \approx 84$--$470$.

On recent consumer GPUs, the double-precision floating-point throughput is
typically $1/32$ or $1/64$ of the single-precision and 32-bit integer throughput,
which at first glance appears close to the required ratio of 65.
However, since the FFT and NTT are fundamentally memory-bound algorithms,
the gap in arithmetic throughput does not translate directly into
a proportional gap in FFT and NTT execution time,
making \cref{eq:speedup-condition} difficult to satisfy even on such hardware.

Although this paper focuses on power-of-two-length FFTs, the Bluestein FFT is applicable to arbitrary lengths, so the proposed method naturally extends to arbitrary-length FFTs.
For an arbitrary-length Bluestein FFT, even without the proposed method, three power-of-two-length FFT calls are required.
This effectively introduces a factor of $1/3$ on the right-hand side of \cref{eq:speedup-condition}.
Substituting the values from our experiments, the required ratio $T_{\mathrm{base}} / T_{\mathrm{NTT}}$ reduces to approximately 21.
Therefore, for arbitrary-length FFTs, achieving competitive execution time with the proposed method may be more realistic.

Recently, Ozaki et al.~\citeyearpar{Ozaki2025} proposed a variant of the Ozaki scheme that uses the CRT in the splitting phase, reducing the number of split component products from $O(k^2)$ to $O(k)$, where $k$ is the number of splits.
If this technique can be extended to cyclic convolution, it could further reduce the number of NTT calls in our method.

A key advantage of the Ozaki scheme for matrix-matrix products is its portability:
it relies on BLAS, a standardized and widely available library
with vendor-optimized implementations at various precisions on virtually every platform.
In contrast, the proposed method requires an NTT as its core building block.
Unlike matrix multiplication and the FFT, for which standardized libraries such as BLAS and FFTW are widely available,
there is no comparable NTT library that provides
optimized implementations across various bit widths and architectures.
However, the NTT is widely used as a core operation in lattice-based cryptography
and fully homomorphic encryption, and efficient NTT implementations
are being actively developed.
In particular, several GPU-based implementations have been proposed~\cite{Ozcan2025,Ji2025,Yin2025,Sugizaki2026a}.
Since the proposed method uses the NTT in a modular fashion,
an efficient NTT implementation in a high-throughput precision on a platform with substantial performance gaps across arithmetic precisions
would enable the method to target various precisions
simply by adjusting the number of splits.

A natural approach for exploiting such performance gaps would be to use NTTs at a lower precision directly, such as 16-bit or 8-bit integers, as the Ozaki scheme does for matrix multiplication.
However, the maximum length of an NTT is determined by the size of the modulus, and the correspondingly smaller moduli would severely restrict the achievable NTT length, limiting the practical applicability of the method.
Alternative approaches have recently been proposed by Yin et al.~\citeyearpar{Yin2025} and Sugizaki and Takahashi~\citeyearpar{Sugizaki2026a} and to exploit lower-precision arithmetic units for NTT computation while alleviating the transform-length limitation.
Since our method reduces target-precision FFT computation to NTT calls, combining it with such a technique could potentially enable target-precision FFTs to be computed entirely using lower-precision arithmetic units.

\section{Conclusion} \label{sec:conclusion}
In this paper, we proposed a method for computing FFTs at a target precision using lower-precision FFTs.
The method is based on the Bluestein FFT, which computes the DFT via cyclic convolution, and applies the Ozaki scheme to this cyclic convolution.
When using floating-point FFTs, guaranteeing exact computation of the cyclic convolution would require a large number of FFT calls in the split component convolutions; we therefore used the NTT, an FFT over a finite field, together with the CRT.
We also introduced an upper bound $K$ on the number of splits to control the number of NTT calls, as well as an NTT-domain accumulation strategy that further reduces the number of inverse NTTs required.

We implemented a double-precision Bluestein FFT using 32-bit NTTs and evaluated its accuracy and computational cost.
The proposed method effectively reduced the relative error of the FFT compared with that for existing double-precision FFT implementations, including FFTW and TS-based FFTs.
In particular, the relative error did not increase with the FFT length, suggesting that the Ozaki scheme suppresses the accumulation of rounding errors.
Depending on the distribution of the input data, the proposed double-precision FFT required up to 268 32-bit NTT and inverse NTT calls for FFT lengths from $2^{10}$ to $2^{18}$.
By setting the upper bound on the number of splits to $K = 3$, the number of 32-bit NTT and inverse NTT calls was reduced to at most 96 while maintaining accuracy.
Furthermore, by allowing up to $L = 3$ NTT-domain accumulations before inverse NTT computation, the count could be further reduced to 64, although this may decrease accuracy depending on the input distribution.

On an Intel Xeon Platinum 8468, the measured execution time is approximately 107 times ($n = 2^{15}$) to 1315 times ($n = 2^{10}$) that of FFTW's double-precision FFT.
The 32-bit NTT/NTT$^{-1}$ computations accounted for approximately 80\% of the total execution time.
If the 32-bit NTT/NTT$^{-1}$ were optimized to the level of FFTW's single-precision FFT, their share of the total execution time would drop to 11\% ($n = 2^{10}$) to 62\% ($n = 2^{15}$).
Moreover, since the time ratio between the NTT and non-NTT parts grows only as $O(\log n)$ with respect to the FFT length, optimization of the non-NTT parts is also necessary.

For the proposed method to outperform a baseline double-precision FFT, a single NTT call must be significantly faster than the baseline FFT, which remains difficult on current hardware; however, for arbitrary-length FFTs, the speedup condition is relaxed, making competitive execution time more realistic.
Although there is no standardized NTT library comparable to BLAS or FFTW, the NTT is widely used in lattice-based cryptography and fully homomorphic encryption, and efficient NTT implementations are being actively developed.
In particular, several GPU-based implementations have been proposed.
Since the proposed method uses the NTT in a modular fashion, an efficient NTT implementation in a high-throughput precision on a platform with substantial performance gaps across arithmetic precisions would enable the method to target various precisions simply by adjusting the number of splits.

\bibliographystyle{ACM-Reference-Format}
\bibliography{bib}

\end{document}